\documentclass[usenatbib,babel]{aa}
\usepackage[varg]{txfonts}

\usepackage[english,english]{babel}
\usepackage{amsmath}
\usepackage{amssymb,amsfonts,textcomp}
\usepackage{array}
\usepackage{supertabular}
\usepackage{hhline}
\usepackage{hyperref}
\usepackage[usenames]{color}
\hypersetup{dvips, colorlinks=true, linkcolor=black, citecolor=black, filecolor=black, urlcolor=black}

\def\gtrsim{\lower.5ex\hbox{$\; \buildrel > \over \sim \;$}}
\usepackage{graphicx}

\newcommand{\hagn}{\mbox{{Horizon-AGN\,\,}}}

\def \apjl {{\it ApJ Let.}}
\def \apjs {{\it ApJ Sup.}}

\def \aap {{\it A\&A}}

\definecolor{grey}{rgb}{0.75,0.75,0.75}
\definecolor{Orange}{rgb}{1.0,0.5,0.15}
\definecolor{brown}{rgb}{0.7,0.25,0.0}
\definecolor{pink}{rgb}{1.0,0.5,0.5}
\definecolor{darkerred}{rgb}{0.8,0,0}
\definecolor{darkerblue}{rgb}{0,0,0.8}
\definecolor{Blue}{rgb}{0,0.08,0.65}
\definecolor{Red}{rgb}{0.65,0.08,0.05}
\definecolor{Green}{rgb}{0.15,0.45,0.25}

\begin{document}

\title{Caught in the rhythm: }
\subtitle{how satellites settle into a plane around their central galaxy}

\author{C. Welker 
	\inst{1,2}
	\and
	Y. Dubois \inst{2}
	\and 
	C. Pichon \inst{2,3}
	\and 
	J. Devriendt \inst{4,5}
	\and 
	N. E. Chisari \inst{4}
	}
	
\institute{  ICRAR, The University of Western Australia, Crawley, Perth, 6009, WA, Australia\\
		  \email{charlotte.welker@uwa.edu.au}
		\and
		CNRS and UPMC Univ. Paris 06, UMR 7095, Institut d'Astrophysique de Paris,
                98 bis Boulevard Arago, F-75014 Paris, France
                \and
                Institute of Astronomy, University of Cambridge, Madingley Road, Cambridge, CB3 0HA, United Kingdom
                \and
                Sub-department of Astrophysics, University of Oxford, Keble Road, Oxford, OX1 3RH, United Kingdom
                \and
                Observatoire de Lyon, UMR 5574, 9 avenue Charles Andr\'e, Saint Genis Laval 69561, France
                }

\date{Accepted . Received ; in original form }

\abstract
	{The anisotropic distribution of satellites around the central galaxy of their host halo is both well-documented in observations and predicted by the $\Lambda$CDM model. However its amplitude, direction and possible biases associated to the specific dynamics of such satellite galaxies are still highly debated. }
	 {Using the cosmological hydrodynamics simulation Horizon-AGN, we investigate the spatial distribution of satellite galaxies relative to their central counterpart and to the cosmic web, in the redshift range between 0.3 and 0.8.}
    	{Haloes and galaxies are identified  and their kinematics computed using their dark matter and stellar particles  respectively. Sub-haloes are discarded and galaxies lying within $5\, R_{vir}$ of a given halo are matched to it. The filamentary structure of the cosmic web is extracted from the density field - smoothed over a 3 $h^{-1} \rm Mpc$ typical scale - as a network of contiguous segments. We then investigate the distribution function of relevant angles, most importantly the angle $\alpha$ between the central-to-satellite separation vector and the group's nearest filament, aside with the angle $\theta_1$ between this same separation and the central minor axis. This allows us to explore the correlations between filamentary infall, intra-cluster inspiralling and the resulting distribution of satellites around their central counterpart. }
    { We find that, on average, satellites tend to be located on the galactic plane of the central object. This effect is detected for central galaxies with a stellar mass 
    larger than $10^{10} \, \rm M_{\odot}$ and found to be strongest for red passive galaxies, while blue galaxies exhibit a weaker trend. For galaxies with a minor axis parallel to the direction of the nearest filament, 
    we find that the coplanarity is stronger in the vicinity of the central galaxy, and decreases when moving towards the outskirts of the host halo. 
    By contrast, the spatial distribution of satellite galaxies relative to their closest filament follows the opposite trend: their tendency to align with them dominates at large distances 
    from the central galaxy, and fades away in its vicinity. Relying on mock catalogs of galaxies in that redshift range, we show that massive red centrals with a spin perpendicular
    to their filament also have corotating satellites well aligned with both the galactic plane and the filament. On the other hand, lower-mass blue centrals with a spin parallel 
    to their filament have satellites flowing straight along this filament, and hence orthogonally to their galactic plane. The orbit of these satellites is then progressively bent towards a better 
    alignment with the galactic plane as they penetrate the central region of their host halo.}
    {The kinematics previously described are consistent with satellite infall and spin build-up via quasi-polar flows, followed by a re-orientation of the spin of massive red galaxies through mergers.}


 \titlerunning{Satellites alignment in Horizon-AGN}
 
\maketitle

\section{Introduction}

The complex interactions between central galaxies and their satellites in dark matter haloes have received a great deal of attention from both theorists 
        and observers over the past decades. Numerous questions regarding the precise distribution of satellites around their central counterparts - noticeably their orientation with respect to the 
        central galactic plane and the connection of this latter with the surrounding cosmic web - have been investigated in both observational and numerical works. Although a few observational 
        studies have claimed to observe polar alignment of satellites around centrals (the so-called ``Holmberg effect'', \citealt{Holmberg69,Zaritsky97}) or none at all \citep{Hawley75,Phillips15}, 
        most of of these studies show that satellite galaxies are  distributed within the galactic plane of their central galaxy~\citep{Brainerd05,Yang06,Sales09,Wang10,Nierenberg12, smithetal15}, and possibly 
        corotating with it~\citep{Ibata13}. Those findings are in fair agreement with predictions from $N$-body \citep{Aubert04, Zentner05} and hydrodynamical cosmological simulations \citep{Dong14}, 
        although it is still debated whether these models predict planes of satellites as thin as those observed around  the Milky-Way and Andromeda~\citep{Bahl14, Pawlowskietal14, Gilletetal15}. 
        The analysis of galaxy groups in the Sloan Digital Sky Survey (SDSS, \citealt{Abazajian09}) has confirmed this trend and established that the signal is stronger for massive red central galaxies, 
        especially in the inner regions of the halo \citep{Yang06}. 

	The interest in the distribution of satellites stems in part from their expected ability to trace the underlying dark matter density of their host halo. In the standard picture, progenitors of satellite galaxies end up 
	orbiting around their host with a distribution tracing its potential well, hence the geometry of the underling dark matter density. A robust consequence of this scenario \citep{Wang05,Agustsson10} is therefore  
	that the perceived concentration of satellites in the rotation plane of their host halo is a direct tracer of its triaxiality \citep{BarnesE87,Warren92,Yoshida2000,Meneghetti01,Jing02} inherited from its hierarchical 
	build-up within an anisotropic cosmic web. Moreover, there is evidence that alignment trends of brightest cluster galaxies, luminous
	red galaxies and groups/clusters are also preserved on large scales (up to $100\,  h^{-1}\, \rm Mpc$ \citealp{Binggeli82,Plionis02,Hopkins05,Mandelbaum06,Hirata07,Okumura09,Niederste10,Joachimi11,Paz11,Smargon12,Li13,Singh14}) 
	as a consequence of hierarchical structure formation, stretching and re-orientation from tidal interactions \citep{Croft00,Lee00,Crittenden01,Catelan01,Hirata04b,Hirata10,Blazek11,Blazek15,Schaefer15,codis15b}.
	
	The anisotropic distribution of satellites is thus a prediction of the $\Lambda$CDM model but its amplitude and the possible additional biases arising from the specific dynamics of satellite galaxies are still highly debated. 
	Numerical simulations have in particular pointed out a possible degeneracy  regarding the anisotropic distribution of satellites arising from the effect of halo ellipticity and the ongoing - unrelaxed -  anisotropic cosmic infall  \citep{Aubert04,Pichon11}. 
	The population of massive halos have a spin preferentially orthogonal to their host filament~\citep{vanhaarlem93,Tormen97,Bailin08,Paz11,Codis12,Zhang13} since they are the product of multiple mergers between pairs of objects drifting along the cosmic web,
	hence with orbital angular momentum is preferentially perpendicular to filaments along which they flow~\citep{duboisetal14, Welker14}.

	As a consequence, the elongation of the halo and filamentary infall share a unique direction. The tendency of satellites to orbit in the central galactic plane is therefore not a mere tracer of the halo triaxiality, 
	but is also naturally enhanced by the continuing infall of satellites \citep{Aubert04,Knebe04,Wang05,Zentner05}. Recent observations of planes of satellites for M31 or the Milky Way~\citep{Ibata13, Libeskind15} 
	and the discovery of alignment trends in the SDSS by \cite{Paz08} \citep[see also][]{Tempel15} strongly support this claim. 	
	
	However, it is still unclear how strongly sustained anisotropic cosmic infall of satellites impacts their observed angular distribution compared to the possibly more relaxed diffuse dark matter component. It may however be important to make a distinction between a  dynamical angular bias (sustained polar accretion of satellites, gravitational torques from the central disc and halo) specific to the dense baryonic substructures, and a purely geometrical angular anisotropy related to the elongation of the halo and affecting the diffuse dark matter component likewise.
	Getting a better understanding of how satellite galaxies flow along the host filament and sink into their host halo, how this process depends on the orientations of infall relative to that of the central galaxy, and how it affects their statistical 
	distribution, is, thus, of major importance.
		
	Beyond the goal of understanding galactic accretion, intrinsic alignments of galaxy shapes are widely regarded as a contaminant to weak gravitational lensing measurements
	\citep{Croft00,Crittenden01,Catelan01,Bernstein02,Hirata04,Mandelbaum06,Hui08,Schneider10,Joachimi10}. They could play a particularly important role in upcoming cosmic shear measurements, potentially biasing constraints 
	on the evolution of  the dark energy equation of state \citep{Kirk10,Krause15}. The need to access information  on the nonlinear scales of the cosmic shear power spectrum  to constrain dark energy makes numerical hydrodynamical 
	simulations  useful  to study the mechanisms that lead to alignments \citep{Tenneti14b,Codis15,Velliscig15,Chisari15}, to build a halo model to describe them \citep{Schneider10} and to constrain alignment bias parameters \citep{Blazek15}. 	
	Indeed, the coplanarity of satellites in the vicinity of a central massive galaxy could lead to an alignment signal that would contaminate such lensing measurements. It would induce a correlation between the shape of the central and 
	the location of the satellites. In particular, coherent alignments of galaxies with the filaments that define the large-scale structure of the Universe could also produce a contamination to cosmic shear and galaxy-galaxy lensing. 
	A complementary analysis to the work presented in this paper is given in \citet{Chisari15}, which relates the shapes of galaxies in the simulation and their correlations to currently available models for intrinsic alignments. 
	In the present study, we analyse the distribution of galaxy satellites around their central galaxy in the cosmological hydrodynamics simulation Horizon-AGN~\citep{duboisetal14} primarily between redshift $z=0.8$ and $z=0.3$, and how this
	distribution is reconfigured as satellites approach the central galaxy.
	
        This paper is structured as follows: after a short review of the numerical setup and methods used in Section \ref{section:virtual}, Section \ref{section:orientation} focuses on the description of alignment trends measured in Horizon-AGN. 
        In Section \ref{section:dynamics}, we present a more detailed kinematic analysis of the corotation features found for satellites of massive halos at various stages of evolution. Mock observations are presented in  \ref{section:mock}, 
        where we further study the impact of various parameters such as the shape of the central galaxy on the alignment trends. Finally, Section \ref{section:discussion} discusses these results in the scope of
        recent observations and Section~\ref{section:conclusion} summarizes our main results.

\section{Numerical methods \& definitions}
\label{section:virtual}

Let us first review briefly the numerical setup and methods used to produce virtual data sets and analyze galactic orientation.
\subsection{The Horizon-AGN simulation}
\label{section:HAGN}

The details of the Horizon-AGN\footnote{\url{www.horizon-simulation.org}} simulation, that we only briefly describe here, can be found in~\cite{duboisetal14}. The \hagn\!\! simulation is run in a 
$L_{\rm box} = 100\,  h^{-1} \rm Mpc $ cube with a $\Lambda$CDM cosmology with total matter density $\Omega_{\rm  m}=0.272$, dark energy density $\Omega_\Lambda=0.728$, amplitude of the matter power spectrum $\sigma_8=0.81$, baryon density $\Omega_{\rm  b}=0.045$, Hubble constant $H_0=70.4 \, \rm km\,s^{-1}\,Mpc^{-1}$, and $n_s=0.967$ compatible with the WMAP-7 data~\citep{komatsuetal11}. 
The total volume contains $1024^3$ dark matter (DM) particles, corresponding to a DM mass resolution of $M_{\rm  DM, res}=8\times10^7 \, \rm M_\odot$, and initial gas resolution of $M_{\rm gas,res}=1\times 10^7 \, \rm M_\odot$. It is run with the {\sc ramses} code~\citep{teyssier02}, and the initially coarse $1024^3$ grid  is adaptively refined down to $\Delta x=1$ proper kpc, with refinement triggered in a quasi-Lagrangian manner: if the number of DM particles in a cell becomes greater than 8, or if the total baryonic mass reaches 8 times the initial baryonic mass resolution in a cell. It results in a typical number of $7\times 10^9$ gas resolution elements (leaf cells) in the \hagn\!\! simulation at $z=0$.

Heating of the gas from a uniform UV background takes place after redshift $z_{\rm  reion} = 10$ following~\cite{haardt96}. Gas can cool down to $10^4\, \rm K$ through H and He collisions with a contribution from metals using rates tabulated by~\cite{sutherland93}. 
Star formation occurs in regions of gas number density above $n_0=0.1\, \rm H\, cm^{-3}$ following a Schmidt law: $\dot \rho_*= \epsilon_* {\rho_{\rm g} / t_{\rm  ff}}$,  where $\dot \rho_*$ is the star formation rate mass density, $\rho_{\rm g}$ the gas mass density, $\epsilon_*=0.02$ the constant star formation efficiency, and $t_{\rm  ff}$ the local free-fall time of the gas. Feedback from stellar winds, supernovae type Ia and type II are included into the simulation with mass, energy and metal release, assuming a Salpeter initial mass function.
The formation of black holes (BHs) is also taken into account. They can grow by gas accretion at a Bondi-capped-at-Eddington rate and coalesce when they form a tight enough binary. BHs release energy in a heating or jet mode (respectively ``quasar'' and ``radio'' mode) when the accretion rate is respectively above and below one per cent of Eddington, with efficiencies tuned to match the BH-galaxy scaling relations at $z=0$~\citep[see][for details]{duboisetal12}.

\subsection{Galaxies and haloes: detection and matching}
\label{section:structures}

Galaxies are identified using the most massive sub-node method~\citep{tweedetal09} of the AdaptaHOP halo finder~\citep{Aubert04} operating on the distribution of star particles with the same parameters than in~\cite{duboisetal14}. 
Unless specified otherwise, only structures with a minimum of $N_{\rm  min}=50$ star particles are considered, which typically selects objects with masses larger than  $1.7 \times 10^8 \, \rm M_\odot$. Catalogues containing up to $\sim 350 \, 000$ haloes and $\sim 180 \, 000$ galaxies are produced for each redshift output analysed in this paper ($0.3 < z < 0.8$). This study focuses on a rather low redshift range $z=0.3-0.8$, accessible to current observations. This detection threshold in terms of number of particles might seem low with respect to possible lack of convergence of shape tracers for poorly defined centrals, however, we apply a higher threshold to select central galaxies afterwards: galaxies with a stellar mass $M_{\rm g}< 10^9 \, \rm M_\odot$ are systematically excluded from the central galaxy sample. Unless specified otherwise, results are stacked over the whole range of redshifts $0.3<z<0.8$ (all pairs found are stacked).

        To match central galaxies with their host halo, we associate each galaxy with the halo located at the shortest distance from it (measured as the distance between their respective centres of mass). In case of 
	multiple galaxies associated with the same halo, we identify the central galaxy as the most massive galaxy contained within a sphere of radius $R=0.5 \, R_{\rm vir}$ with $R_{\rm vir}$ the virial radius 
	of the halo. Unless otherwise specified, sub-haloes are not considered as able to host central galaxies. The satellite galaxies of a halo are defined as all the galaxies -- excluding the central --  situated within a sphere of radius 
	$R=5 \, R_{\rm vir}$. This includes galaxies belonging to sub-haloes of the host halo, but also to sub-haloes of neighbouring haloes. This seemingly too large a scale is chosen so as to ensure that incomplete 
	randomisation of walls and grid periodicity on the largest scales of the simulated volume do not affect the 3D angular statistics presented in our analysis.
	Further cuts in distance to the central galaxy can be performed afterwards and will be specified in each case.

	At $z=0.3$, approximately $16\,000$ main haloes are inhabited. The richest one hosts 678 galaxies; 263 haloes contain more than 30 galaxies; $2\,220$, haloes more than 10; and $6\,622$ host at least 2 galaxies. 
	This distribution does not show strong variations across the redshift range analysed in this work.

\subsection{Synthetic colors}
\label{section:syncol}

For each galaxy, the absolute AB magnitudes and rest-frame colours are computed using single stellar population models from~\cite{Bruzual03} assuming a Salpeter initial mass function. Each star particle contributes to a flux per frequency that depends on its mass, age, and metallicity. Contributions to the reddening of spectra from internal (interstellar medium) or external (intergalactic medium) dust extinction are not taken into account. Computing the spectrum
	of all star particles and convolving with the $u$, $g$, $r$, and $i$ filters from the SDSS, we then build two-dimensional projected maps of each galaxy (satellites are excised using the galaxy finder) and compute their luminosities
	in these wavebands. 

\subsection{Galaxy morphologies and kinematics}
\label{section:kin}

	We compute stellar particle kinematics for all central galaxies and satellites in the sample.
	The angular momentum -- or spin -- of a galaxy is defined as the total angular momentum of the star particles it contains and is measured with respect to the densest star particle (which proves a more robust estimator of the galaxy
	centre than its centre of mass in cases were the central galaxy is in the process of merging):
	\begin{eqnarray}
	\mathbf{L_{\rm g}}=\sum_{i} m_{i}(\mathbf{r_{i}} -\mathbf{r_{\rm g}} ) \times (\mathbf{v_{i}} -\mathbf{v_{\rm g}} )\, ,
	\end{eqnarray}
	with $\mathbf{r_{i}}$, $m_{i}$ and $\mathbf{v_{i}} $ the position, mass and velocity of particle {\it i}, $\mathbf{r_{\rm g}}$ the position of the centre of the galaxy
	and $\mathbf{v_{\rm g}}$ its centre of mass velocity. The notation $\mathbf{L_{\rm g}}$ is hereafter used for the angular momentum of the central galaxy while 
	$\mathbf{l_{\rm sat}}^{i}$ denotes the intrinsic angular momentum of its i$^{th}$ satellite.
  
	The {\it separation vector}, or position vector of each satellite in the rest frame of its central galaxy is defined as $\mathbf{r_{\rm gs}}= \mathbf{r}_{s}-\mathbf{r_{\rm g}}$ with $\mathbf{r}_{s}$ the position of the satellite. Its norm is  
	the separation $R_{\rm gs} = || \mathbf{r}_{\rm gs} || $.
 
	 The inertia tensor of each galaxy is computed from the star particle masses ($m^{l}$) and positions ($x^{l}$) (in the barycentric coordinate system of the galaxy):
 	\begin{eqnarray}
 	I_{ij} = \sum_{l} m^{l}(\delta_{ij} (\sum_{k}x^{l}_{k} x^{l}_{k})-x^{l}_{i} x^{l}_{j})\, ,
 	\end{eqnarray}
	where $\delta_{ij}$ is the Kronecker symbol. This inertia tensor is diagonalised, with its eigenvalues $\lambda_{1} > \lambda_{2} > \lambda_{3}$ the moments of the tensor relative to the basis of principal axes $e_{1}$, $e_{2}$ and $e_{3}$.
	The lengths of the semi-principal axes (with $a_{1}<a_{2}<a_{3}$) are derived from the moments of inertia: 
	\begin{eqnarray}
	a_{1}= (5/M_{\rm g})\sqrt{\lambda_{3}+\lambda_{2}-\lambda_{1}} \, , \: \rm along \: e_{1} \nonumber \, , \\
	a_{2}= (5/M_{\rm g})\sqrt{\lambda_{1}+\lambda_{3}-\lambda_{2}} \, , \: \rm along \: e_{2} \nonumber \, , \\
	a_{3} = (5/M_{\rm g})\sqrt{\lambda_{1}+\lambda_{2}-\lambda_{3}} \, , \: \rm along \: e_{3} \nonumber \, .
	\end{eqnarray}
	This allows for an easy estimation of the galactic shape using the tri-axiality ratio $\tau = {(a_{2}-a_{1}) / (a_{3}-a_{2}) }$. Oblate structures (disk-shaped) have $\tau > 1$ while prolate structures (cigar-shaped) have $\tau < 1$.

	For comparison with observations, we also define the corresponding projected quantities along the x-axis of the grid (labeled "X"). Definitions are similar for the positions and inertia tensor with summations restricted to the projected 
	coordinates $(y,z)$. This leads to the eigenvalues $\lambda_{1}^{X}$ and $\lambda_{2}^{X}$, from which we derive the axis $a_{1}^{X} <a_{2}^{X}$.

	As we are interested in studying the corotation  of satellites, i.e. their tendency to align their orbital momentum with the spin of the central galaxy and synchronise their rotation with that of the central disk,
	we also compute the total orbital angular momentum of satellite systems as $\mathbf{L}_{\rm sat}^{\rm orb}$ with similar definitions as previously used to derive spins, but applied to the velocities and positions of satellites.
	Average circular velocities (for the central galaxy or its orbiting system of satellites) are defined as	
 	\begin{eqnarray}
  	\mathbf{v_{\rm rot}} = {{\sum_{i} m_{i}(\mathbf{r_{i}} \times \mathbf{v_{i}} )}\over{ \Sigma_{i} m_{i} r_{i}} }\, ,
  	\end{eqnarray}
  	with $m_{i}$, $\mathbf{r_{i}}$ and $\mathbf{v_{i}}$ the masses, radius and velocities of structures considered: star particles for the central galaxy, satellites for a system of satellites.

	 We will use the notation  $\mathbf{v_{\rm rot}^{\rm gal}}$ for the average rotational velocity of the stellar material within the central galaxy, and  $\mathbf{v_{\rm orb}^{\rm sat}}$ for the average orbital rotational 
	 velocity of the system of satellites.

\begin{figure}
\center \includegraphics[width=0.7\columnwidth]{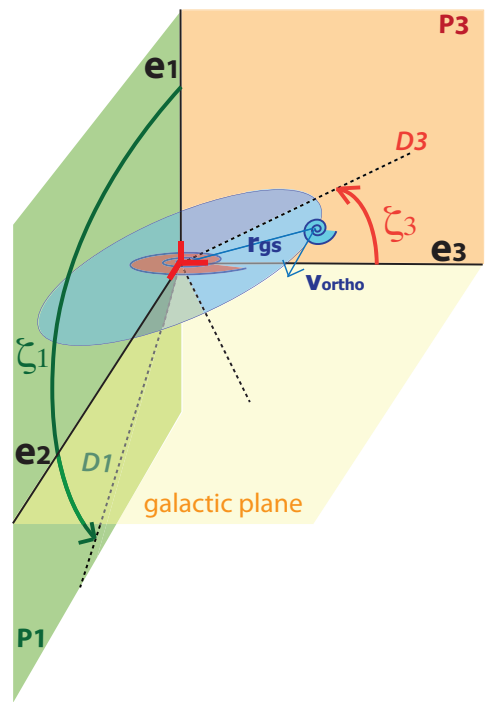}
  \caption{Sketch representation of the angles $\zeta_{1}$ and $\zeta_{3}$ used to describe the orientation of the orbital plane of a satellite (in blue) around its central galaxy (in red), that are used to compute the average orbital plane of satellites and their dispersion ratio (see the text for details). Dashed lines illustrate the intersections between the central galaxy principal planes and the satellite orbital plane. }
\label{fig:orbit}
\end{figure}

	Each satellite has an individual orbital plane defined by $\mathbf{e_{\rho}}=\mathbf{r}_{\rm gs}/R_{\rm gs}$, the direction to the central, and $\mathbf{e}_{\theta}={\mathbf{v}_{\rm ortho}}/{||\mathbf{v}_{\rm ortho}||}$, the direction of the 
	component of its velocity orthogonal to $\mathbf{r}_{\rm gs}$. The intersections, $D_{1}$ and $D_{3}$, of such a plane with the planes $\mathbf{P_{1}}=(\mathbf{e_{1}},\mathbf{e_{2}})$ and  $\mathbf{P_{3}}=(\mathbf{e_{1}},\mathbf{e_{3}})$ of the central galaxy allow 
	us to compute two orientation angles $\zeta_{1}$ between $D_{1}$ and $\mathbf{e_{1}}$, and $\zeta_{3}$ between $D_{3}$ and $\mathbf{e_{3}}$, respectively. An illustration of these angles can be found in Fig.~\ref{fig:orbit}.
	
	  Averaging angles $\zeta_{1}$ and $\zeta_{3}$ for all the satellites in the 
	system, we obtain two angles that define their mean orbital plane. This allows us to compute the {\it dispersion ratio}:  $\sigma_{\rm plane}/||\mathbf{L}_{\rm sat}^{\rm orb}||$ with 
	$\sigma_{\rm plane}=\sum_{i} ||\mathbf{j}_{\rm sat}^{\rm orb}||_{i}$ and $||\mathbf{j}_{\rm sat}^{\rm orb}||_{i}$  the norm of the projected orbital momentum of satellite {\it i} on the mean orbital plane of all satellites. 
	This measures the dispersion of the orientation of the angular momentum of the satellites around the mean rotation plane. This parameter drops to zero if satellites are all rotating in the same plane. Similarly, we 
	measure the {\it corotation ratio}: ${|L_{zc}|}/{||\mathbf{L}_{\rm sat}^{\rm orb}||}$, where $L_{zc}$ is the projection of the total orbital momentum on the spin axis of the central galaxy. 
        
\begin{table*}
{ \center
\begin{tabular}{|l|c|r|}
\hline
{\bf Angle} & {\bf Measured quantity}/{\bf definition} \\
 \hline
 Satellite separation vector-filament & $\nu=\cos\alpha$ \\
 \hline
Satellite separation vector-central spin & $\mu=\cos \theta$ \\
\hline
Satellite separation vector-central minor axis & $\mu_{1}=\cos\theta_{1}$\\
\hline
Satellite separation vector-central major axis & $\mu_{3}=\cos\theta_{3}$\\
\hline
Central spin-filament & $\nu_{\rm g}=\cos\alpha_{\rm g}$\\
\hline
Central minor axis-filament & $\nu_{\rm g1}=\cos\alpha_{\rm g1}$\\
\hline
Projected (2D): satellite separation vector-central major axis & $\theta_{\rm x}$\\
\hline
Projected (2D): satellite separation vector-filament & $\alpha_{\rm x}$\\
\hline
Satellites total orbital momentum-central spin &  $\cos\phi$\\
\hline
Satellite spin-central spin & $\cos\chi$\\
\hline
Satellite minor axis-satellite separation vector & $\cos\chi_{1}$\\
\hline
Satellite spin-satellite separation vector & $\cos\chi_{\rm s}$\\
\hline
\end{tabular}
\center\caption{Definitions of the different angles used in this work; see also Fig.~\ref{fig:sketch3}.}
\label{angles}
}
\end{table*}

Table~\ref{angles} summarises the definitions of all the angles used in this paper to follow alignment trends and the amount of corotation of satellites relative to their central galaxy.

\subsection{Characterization of the cosmic web}

In order to quantify the orientation of galaxies relative to the cosmic web, we use a geometric three-dimensional ridge extractor called the {\sc skeleton}~\citep{sousbie09} computed from the full volume DM density distribution 
	sampled on a $512^3$ cartesian grid. This density distribution is smoothed with a gaussian kernel of length $3 \, h^{-1}$ comoving $\rm Mpc$. The orientation and distribution of galaxies can be measured relative to the direction of 
	the closest filament segment. Note that such filaments are defined as ridge lines of the density field and therefore have no thickness. The closest filament of a given galaxy is thus simply the segment whose distance to the galaxy 
	is the shortest. However, all central galaxies in the sample are separated from their nearest filament by less than 1 Mpc, and the vast majority of them by less than 0.5 Mpc (the peak of the galaxy distance-to-filament 
	distribution lies around 0.2 Mpc).

\section{Orientation of satellites}
\label{section:orientation}

We investigate the alignment  of satellites galaxies in the redshift range $0.3<z<0.8$.
We characterise two distinct trends: 
\begin{itemize}
\item The tendency of satellites to lie on the galactic plane of their central galaxy: the {\it coplanar trend}. This can be analysed computing either $ \mu=\cos \theta$, the cosine of the angle between the satellite's position vector and the spin of the central, or $ \mu_{1}=\cos \theta_{1}$, with the minor axis of the central (see Fig.~\ref{fig:sketch3}). In this paper, we mostly  focus on the latter, because it is more closely related to observational methods (see \citealp{Yang06,Tempel15}). 
\item The tendency of satellites to align within the nearest filament: the {\it filamentary trend}. To quantify it, we compute $ \nu=\cos \alpha$ the cosine of the angle between the satellite's position vector and the direction of the closest filament. The {\it filamentary trend} is detected when there is an excess probability $\xi>0$ for $\cos\alpha=1$.
\end{itemize}

\subsection{Mass dependency of in-plane alignments}
\label{section:mass-align}

\subsubsection{Results}

\begin{figure}
\center \includegraphics[width=0.9\columnwidth]{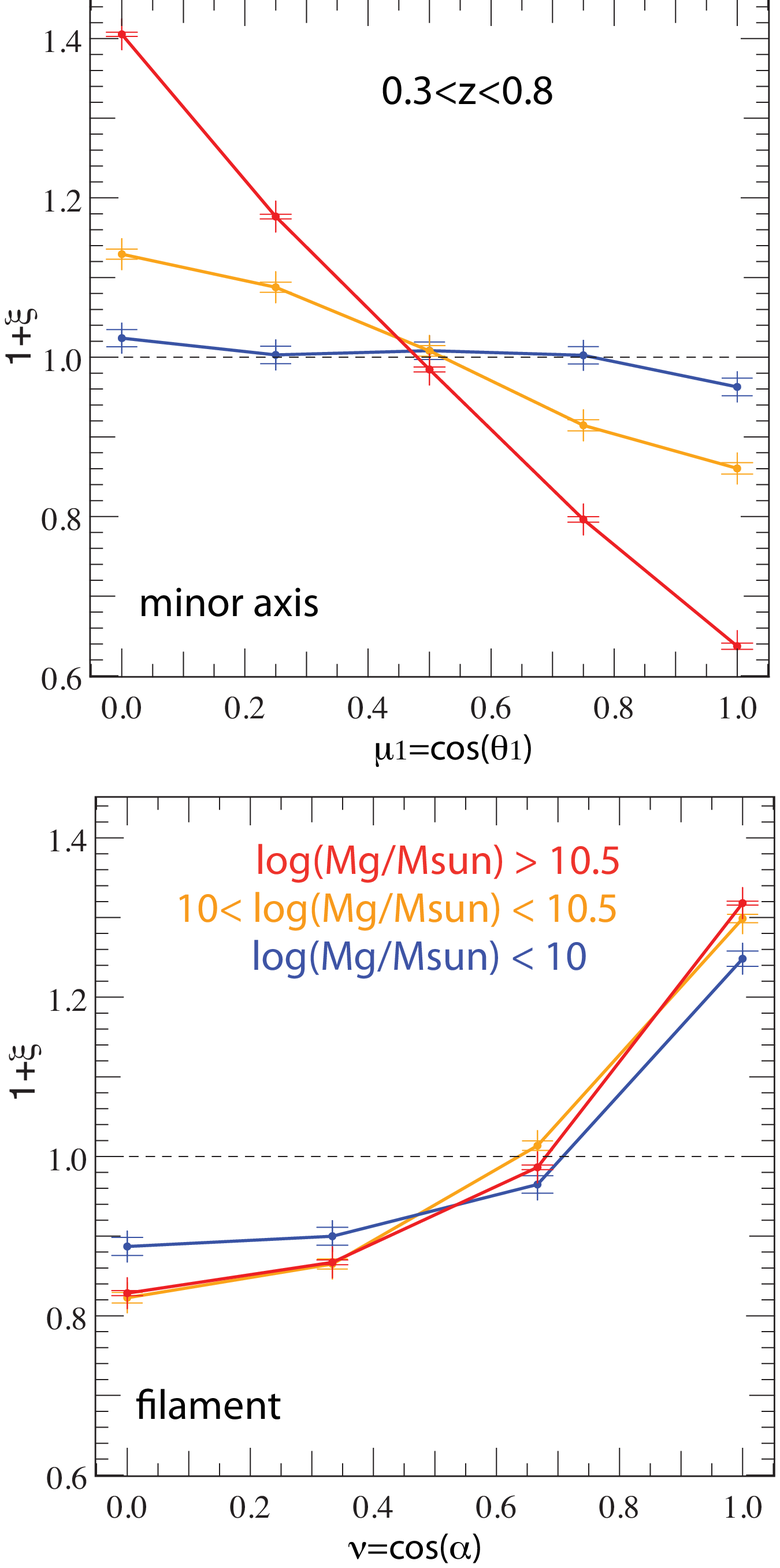}
  \caption{{\sl Top panel:} excess PDF $\xi$ of $ \mu_{1}=\cos \theta_{1}$, the angle between the minor axis of the central galaxy and the direction towards the center of mass of its satellites, for different central galaxy mass bins, between redshift $0.3<z<0.8$.  Satellites tend to be distributed on the galactic plane of the central and this trend is stronger with the increasing mass of the central. {\sl Bottom panel:} excess PDF of $\nu=\cos \alpha$, the cosine of the angle between the direction of the central's nearest filament axis and the direction towards the center of mass of its satellites. Satellites tend to be strongly distributed along filaments.
  }
\label{fig:mass0}
\end{figure}

Let us first study the {\it coplanar trend}. The top panel of Fig.~\ref{fig:mass0} shows the probability density function (or PDF$\equiv 1+\xi$, with $\xi$ the excess of probability) of $ \mu_{1}=\cos\theta_{1}$,  where $\theta_{1}$ is the unsigned angle in $[0,\pi/2]$ between the minor axis of the central galaxy and $\mathbf{r}_{\rm gs}$ for central galaxies in different mass bins.
	As previously mentioned, we stack results from the satellite distribution around central galaxies over six \hagn outputs in the redshift range $0.3<z<0.8$ equally spaced in redshift and consider only satellites within $5 R_{\rm vir}$ 
	of the central galaxy. The main effect of this stacking is a smoothing of the signal. It was checked that results for individual snapshots are fully consistent with the stacked results, although with larger error bars.

	On average, satellites have a tendency to lie on the galactic plane of the central galaxy, or equivalently, their direction is preferentially perpendicular to the minor axis of the galaxy as shown by the excess PDF, $\xi$, at 
	$\mu_1=0$~\citep[see][for DM sub-halos]{Aubert04}.  However, this conclusion does not hold  equally for all central galaxies: while the distribution of satellites around their central is mostly random for low-mass centrals with 
	$10^{9} < M_{\rm g} <10^{10} \, \rm M_{\odot}$, the alignment strengthens as the central mass increases. For the most massive centrals, galaxy satellites clearly tend to be preferentially distributed in the plane of the central galaxy.
	For central galaxies with stellar mass $M_{\rm g} > 10^{10.5} \, \rm M_{\odot}$, the excess PDF at $\mu_1=0$ is $\xi=40\%$, with $52\%$ of satellites lying outside of a $66^\circ$  ($\cos\theta_{1}<0.4$) double cone 
	of axis perpendicular to the galactic plane, as opposed to $40\%$ for the uniform PDF (dashed line). 
	For intermediate central masses ($10^{10} < M_{\rm g} < 10^{10.5} \, \rm M_{\odot}$), the excess PDF at $\mu_1=0$ is $\xi=13\%$, and $45\%$ of satellites lie outside of the $66^\circ$ inverted bicone.
	No substantial excess is found for lower masses ($\xi<2\%$ at $\mu_1=0$).
	Thus the tendency of satellites to lie on the galactic plane of their central galaxy is directly correlated to this latter's mass. 

	Let us now investigate the tendency of satellites to be distributed along filaments.
	In the bottom panel of Fig.~\ref{fig:mass0}, we show the excess PDF of $\nu=\cos \alpha$, the angle between the direction of the central galaxy nearest filament axis and $\mathbf{r}_{{gs}}$.
	Satellite galaxies tend to align with the direction of the closest filament associated with the central galaxy at the level of $\xi\simeq 30\%$, mostly independently of the central galaxy mass 
	(though we see a slight increase of $\xi$ with $M_{\rm g}$). Here
	$27\%$ of the satellites are contained within a $37^\circ$ cone with axis parallel to the filament ($\cos\alpha>0.8$), as opposed to $20\%$ for angles uniformly distributed on the sphere.
	This effect holds even for the smallest central masses with a decrease in amplitude of less than $1\%$.

\subsubsection{Interpretation}

	Satellites live preferentially in the nearest filament, which suggests that their tendency to align within the galactic plane may be linked to whether the central galaxy is also aligned with the filament or not.
	We can connect this trend to the findings of \cite{Codis12,duboisetal14,Welker14, Laigle15}. These papers found that massive centrals - formed through mergers -  display a spin orthogonal to their nearest filament.
	As a consequence, their galactic plane will be parallel to the filament and we thus expect an excess of satellites in the plane of the central. In other words, this mass-dependent trend for satellites seems directly connected to the   
	already known  mass dependent  spin orientation of centrals.
	However, one should bear in mind that the  scenario developed in the above mentioned papers also advocates that low-mass central galaxies, caught in the winding of the cosmic flows within a filament, are more likely to
	develop a spin parallel to this filament. Hence, one would expect an excess of satellites orthogonal to the galactic plane of these low-mass centrals. 
	At first glance, this does not happen: we find that satellites tend to be randomly distributed around low mass centrals. Several reasons lead to this discrepancy (see Appendix~\ref{section:low-mass-centrals} for detail):
	\begin{itemize}
	\item The spin alignment trend for low mass centrals was actually detected at higher redshift ($z>1$) and was shown to decrease with cosmic time, while the perpendicular orientation trend for higher mass galaxies is expected to strengthen. 
	As a 	consequence, the satellite orientation  on the galactic plane is much weaker around low-mass centrals in our redshift range.
	\item The combined effect of the threshold we use for galaxy detection and of our choice to use the minor axis rather than the spin to quantify alignments on the galactic plane. Indeed, because we only detect structures with	
	 mass above $ 10^{8.5} \, \rm M_{\odot}$, systems of low-mass centrals with satellites in Horizon-AGN are most often pairs of galaxies ($85\%$ of the low-mass sample, or $97\%$ counting three body systems) --the most massive one being labelled 
	 as central-- with a mass ratio close to unity. This implies that the {\it satellite} and the {\it central} are actually interacting galaxies: they mutually affect each other's spin orientation and shape significantly. The shape --as traced by the 
	 minor axis-- is especially impacted by the interaction while the spin is slightly more resilient.
	\end{itemize}
	Consequently, even though low-mass central galaxies tend to display an orientation of spin parallel to the filament on average, this effect is weak and the alignment signal is strongly suppressed for the subsample of these centrals 
	interacting with a close neighbour, and even further suppressed when traced using the minor axis, for satellites in the vicinity of the central galaxy. This latter effect is analysed in greater detail in Appendix~\ref{section:low-mass-centrals},
	 where we show that the alignment of satellites orthogonally to the galactic plane of low mass centrals is actually recovered either using $\mu=\cos \theta$ (central spin-separation angle) rather than $ \mu_{1}=\cos \theta_{1}$, 
	 or focusing on $\mu_{1}=\cos \theta_{1}$ for a sub-sample of more distant satellites and/or lower mass ratio central/satellite pairs.
\begin{figure}
\center \includegraphics[width=0.7\columnwidth]{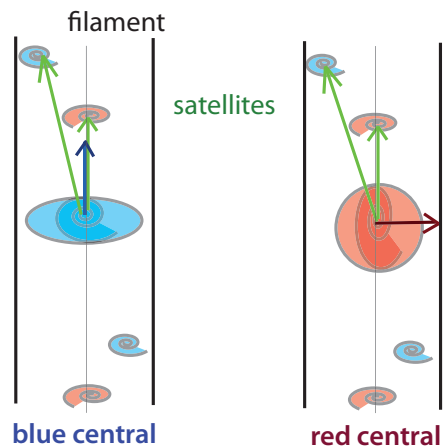}
  \caption{Sketch of the expected results for the {\it filamentary trend}. Satellite galaxies tend to be aligned within the nearest filament. Consequently, they are distributed orthogonally to the spin/minor axis of older red central galaxies, and 
  	more often aligned with the spin/minor axis of young blue central galaxies.}
\label{fig:sketch}
\end{figure}
%


          As a conclusion, after careful analysis these trends seem fully consistent with the statistical measurements of the orientation of the spin of galaxies in the cosmic web highlighted in previous works: low-mass young galaxies fed in vorticity
	 rich regions in the vicinity of filaments~\citep{Laigle15} have their spin parallel to the filament they are embedded in, while older galaxies, more likely to be the products of mergers, are also more likely to display a spin flipped orthogonally 
	 to the filament by a transfer of orbital angular momentum~\citep{duboisetal14, Welker14}. Fig.~\ref{fig:sketch} illustrates this idea that the trends measured in Fig.~\ref{fig:mass0} can therefore be explained by the preferred distribution of satellite galaxies within the filament  closest to the central galaxy. It depicts an "ideal" evolved massive central galaxy sketched in red and a low-mass galaxy in blue.
	 
	Those alignments are  consistent with observational results of  \cite{Yang06} and numerical results of \cite{Dong14} which found a clear alignment of satellites on the galactic plane around red (hence more massive) central galaxies. 
	At the extreme mass end, brightest cluster galaxies are also known to be elongated in the direction of their neighbours; this is the ``Binggeli effect'' \citep{Binggeli82,Niederste10}, and possibly of their closest filament \citep{Paz11}.
	Moreover, our results concerning the filamentary trend are also supported by recent observational studies by \cite{Tempel15} and \cite{Libeskind15} who studied the orientation of plane of satellites in the Local Group and in the SDSS with 
	respect to observationally detected filaments.

\subsection{Satellite alignments versus distance to central}
\label{section:gal-fil}

	While massive central galaxies are overall more likely to show a galactic plane aligned with the nearest filament (or equivalently a spin orthogonal to it), this does not mean however that central galaxies whose spin is aligned with the filament (angle smaller than $37^\circ$ ) constitute a negligible population, even in the higher mass ranges.
	In fact, they still represent $17\%$ of all central galaxies with $M_{\rm g} > 10^{10} \, \rm M_{\odot}$, and moderately misaligned ones 
	(angle between $37^\circ$ and $45^\circ$) account for $23\%$ of the same sample. 

	Such  misalignments between central galaxy planes and nearby filaments imply the existence of galaxy populations for which the {\it filamentary} and {\it coplanar} trends are best described as mutually exclusive (when the spin of the 
	central is well-aligned with the filament) while for "well-behaved" massive centrals (with a galactic plane aligned to the filament) they would lead to qualitatively similar effects. Understanding whether the {\it coplanar trend} can be reduced to 
	being mostly consequence of the {\it filamentary trend} requires further analysis of cases in which both trends might compete.
	To better comprehend where and how this competition occurs, Fig.~\ref{fig:green-orange} focuses on two sub-samples, which we select so as to preserve  statistics:
\begin{figure*}
\center \includegraphics[width=1.8\columnwidth]{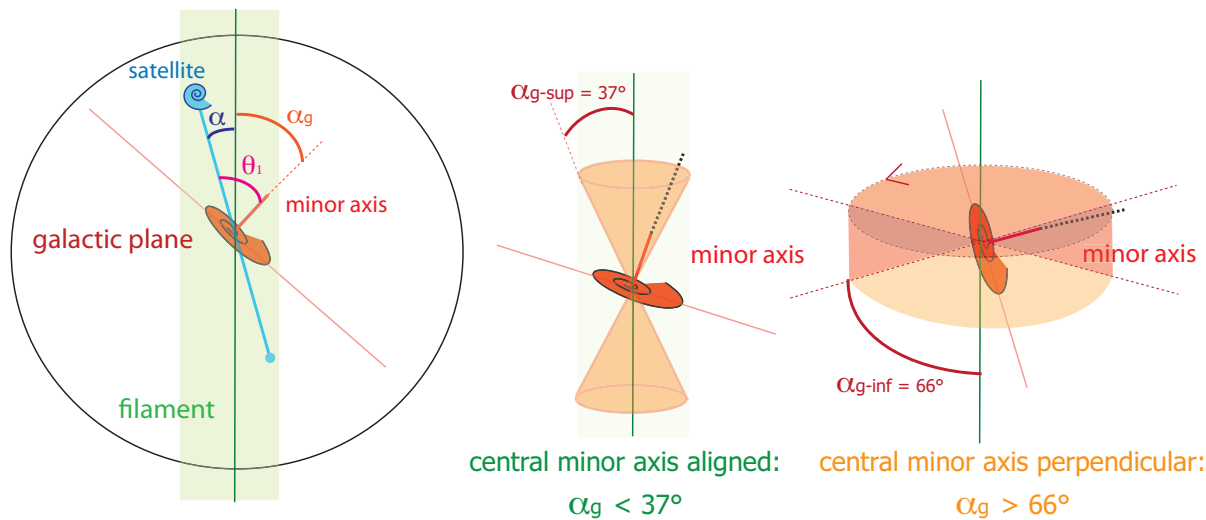}
  \caption{Sketch of the main angles used in our analysis. {\it Left Panel}: definition of $\alpha$, $\theta_{1}$ and $\alpha_{g}$, used to describe the relative orientation of the satellite and the filament, the satellite and the central galaxy, and the central galaxy and the filament respectively. {\it Middle Panel} and {\it Right Panel}: illustration of the solid angle sectors used to define the ``aligned central minor axis" sample (within the $37^{\circ}$ cone) and ``perpendicular central minor axis" sample (outside the $66^{\circ}$ cone) respectively. Note that similar angular sectors can be defined for $\alpha$ and $\theta_{1}$ to describe the orientation of satellites.
  }
\label{fig:sketch3}
\end{figure*}
	\begin{itemize}
	\item central galaxies with minor axis {\sl parallel/aligned} with the filament axis, i.e. their minor axis direction lies within a $37^\circ$ double cone ($\cos \alpha_{\rm g}>0.8$) whose axis of revolution direction coincides with that of the filament, 
	\item central galaxies with minor axis {\sl perpendicular/misaligned} to/with the filament axis, i.e. their minor axis direction lies outside the double cone with opening angle of $66^\circ$ whose axis of revolution direction coincides with that of the filament ($\cos \alpha_{\rm g} <0.4$).
	\end{itemize}
The definitions of these two sub-samples is illustrated on Fig.~\ref{fig:sketch3}.

\subsubsection{Results}

\begin{figure*}
\center \includegraphics[width=1.9\columnwidth]{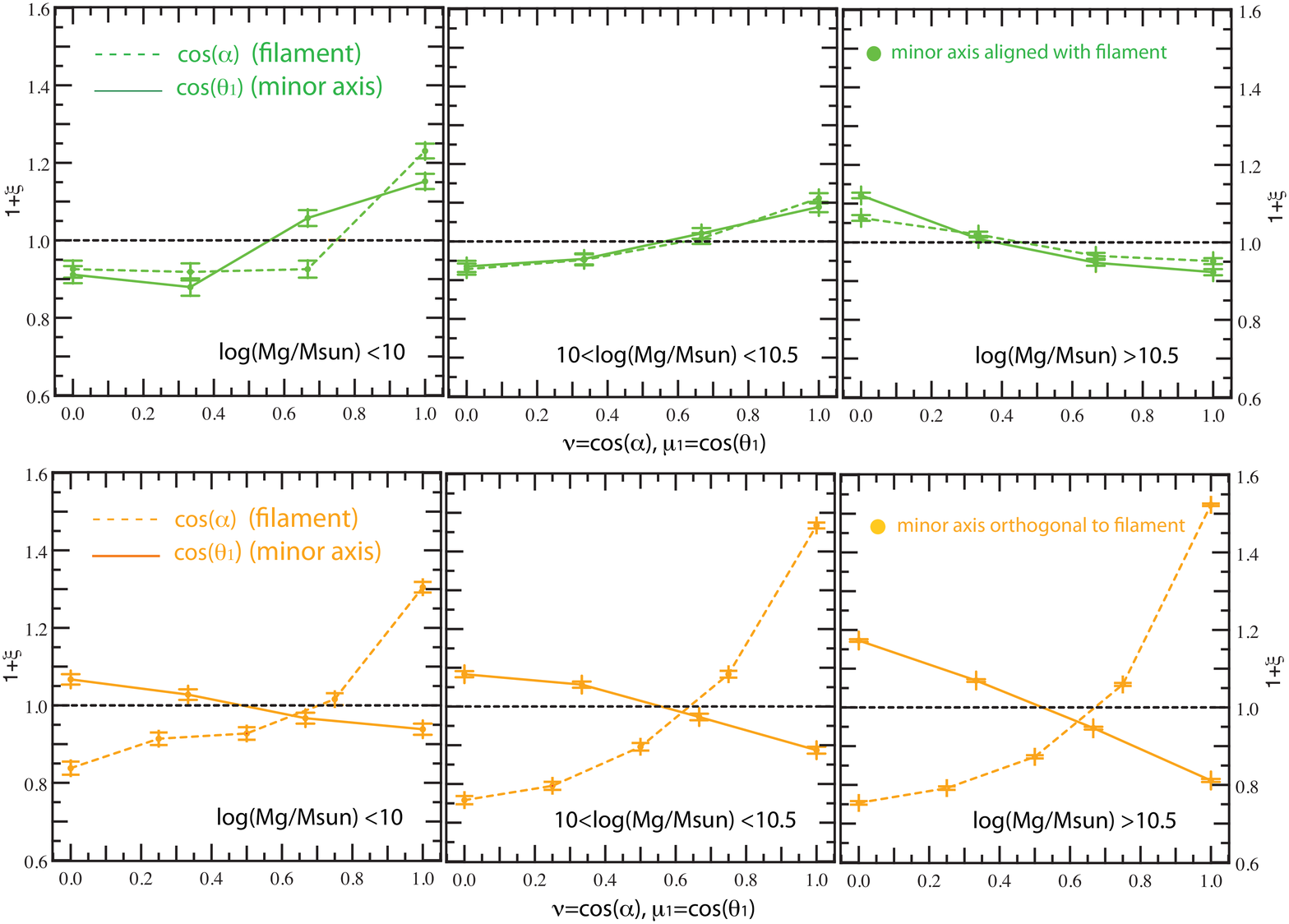}
  \caption{Evolution of the {\it filamentary trend} (dashed line) and the {\it coplanar trend} (solid line) for either the central galaxy's minor axis aligned to the filament axis within a $37^\circ$ cone (top panels), or the central galaxy's minor axis perpendicular to the filament axis within a $37^\circ$ cone (bottom panels). From left to right panel, three different central galaxy mass bins $M_{\rm g}$ are shown as indicated in each panel. Results are stacked for $0.3<z<0.8$. In the top panels, the two trends are mutually exclusive. This reveals a transitional pattern: {\it coplanar trend} takes over for massive centrals while the {\it filamentary trend} is dominant for low mass centrals, consistently with the limited influence of the central torques in this case. In the bottom panels, the two trends reinforce each other. Expectedly, the trends are strengthened for the most massive centrals.
  }
\label{fig:green-orange}
\end{figure*}

	Fig.~\ref{fig:green-orange} shows the excess PDF $\xi$ of $ \mu_{1}=\cos \theta_{1}$ (coplanar trend, solid lines) and $ \nu=\cos \alpha$ (filamentary trend, dashed lines) for both samples and different ranges of central galaxy mass.
 	In the first case (top row), the coplanarity and filamentary trends are mutually exclusive, while in the second case they affect the distribution of satellites in similar ways and become reinforced.  Recall that the {\it coplanar trend} is detected when there is an excess probability $\xi>0$ for $\cos\theta_1=0$, while the {\it filamentary trend} is detected when $\xi>0$ for $\cos\alpha=1$.
	Looking at the first sample of centrals, we see that the coplanar trend dominates for the most massive galaxies ($M_{\rm g} > 10^{10.5} \, \rm M_{\odot}$) with  ${43\%}$ of satellites found on the galactic plane -- i.e. outside the  $66^\circ$ double cone whose axis coincides with the central minor axis-- ($40\%$ for random), even though the filamentary trend has vanished: satellites tend to lie preferentially orthogonally to the nearest filament.
	
	However, the coplanar trend disappears as the central mass decreases ($M_{\rm g} <10^{10.5} \, \rm M_{\odot}$) and as it gets replaced by a polar trend, compatible in this case with the filamentary trend.
	Indeed, the filamentary trend is in contrast recovered for central galaxies with mass $M_{\rm g} <10^{10.5} \, \rm M_{\odot}$, which show a greater degree of alignment to the filament (${24\%}$ in the filament $37^\circ$ double cone instead of $20\%$ for random).
	This investigation reveals that the {\it coplanar trend} is not a mere consequence of the {\it filamentary trend} as they both exhibit complementary transitional patterns with respect to the mass of the central galaxy for centrals of similar spin orientation. 
	This suggests that dissipation in the halo and torques from the central galaxy also strongly impact the orientation of satellites around the most massive centrals, independently from the filamentary infall.
	
Focusing on the second sample of central galaxies whose minor axis is more perpendicular to the filament axis (orange lines), coplanarity and filamentary trends co-exist in all mass bins (excess in $\cos\theta_1=0$ for the solid line, and in $\cos\alpha=1$ for the dashed line).
	Both trends are strengthened for the most massive central galaxies ($M_{\rm g} > 10^{10.5} \, \rm M_{\odot}$), which confirms the distinct role played by galactic plane and filament in orienting satellites.
	The strength of the alignment for this sample are significantly higher than that obtained when the trends compete. The corresponding satellite fractions in the highest mass bin (centrals 
	with masses $M_{\rm g} > 10^{10.5} \, \rm M_{\odot}$) are: ${46\%}$ located on the galactic plane, -- i.e outside the the $66^\circ$ double cone revolving around central the minor axis direction -- 
	and ${32\%}$ located within the $37^\circ$ double cone whose axis is aligned with the filament axis ($20\%$ for random).

In a nutshell, these findings suggest that when the minor axis is orthogonal to the filament, regardless of the mass of the central, satellites always preferentially lie on the galactic plane, which coincides with the direction of the filament. 
	However, when the minor axis is parallel to the filament, the coplanar trend can trump the filamentary trend if the mass of the central galaxy is large enough. It implies that, for these massive galaxies, satellites will preferentially lie on the galactic plane 
	irrespective of the direction of the filament, which in turn suggests that gravitational torques that act on satellite trajectories play a more dominant role in this case.

Quantifying the relative influence of the filament and the joint effect of dissipation angular momentum in the halo and central galaxy torques (which also influence the inner halo shape) is best achieved noticing that these processes operate on different 
	radial scales \citep{danovichetal15}. Far enough from the central galaxy, the filamentary trend should be recovered for all host systems, and we therefore expect a transition from filamentary to coplanar trend as satellites are plunging into 
	haloes hosting misaligned centrals.

\begin{figure*}
\center \includegraphics[width=1.8\columnwidth]{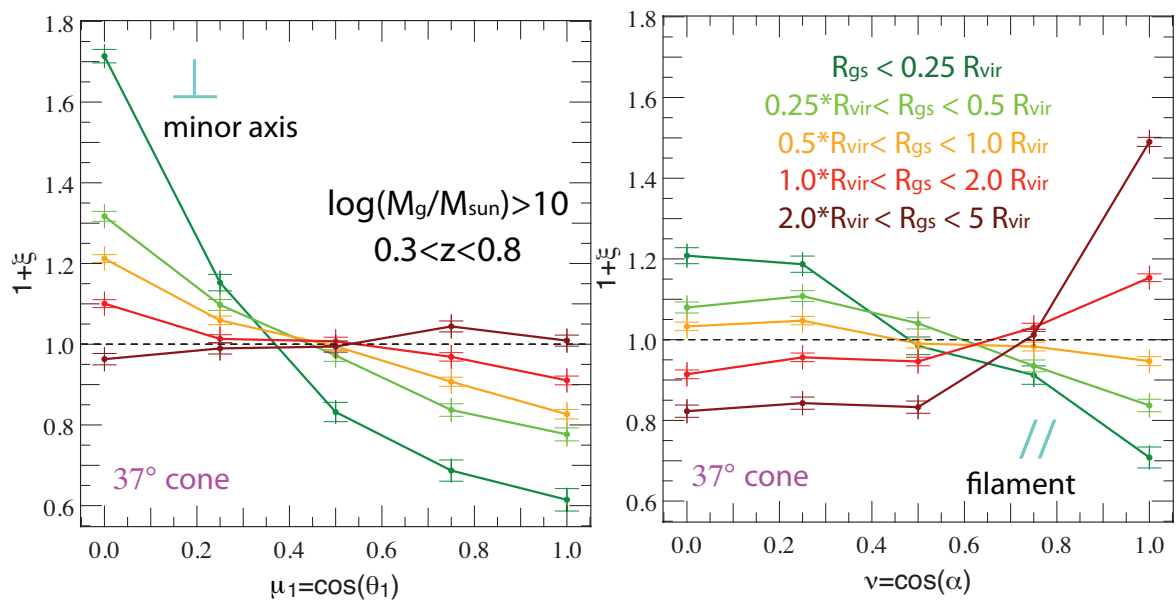}
  \caption{Same as Fig.~\ref{fig:mass0} but binning the sample in distance, $R_{\rm gs}$, from satellite to central, and restricting it to satellites hosted by halos whose central galaxy's minor axis is aligned to the nearest filament within a $37^\circ$ double cone, 
	for which the coplanar and filamentary trends compete. Results are stacked for $0.3<z<0.8$ for different radius $R_{\rm gs} $  bins. 
 	 Satellites close to their central  tend to be distributed on the galactic plane, hence orthogonally to the filament, while satellites in the outskirts of the halo are strongly aligned with the filament and the coplanarity with the central is lost. %
  }
\label{fig:radius_cone}
\end{figure*}

Analysis of the scale segregation inherent to this competition between coplanarity and filamentary trends can be found in Fig.~\ref{fig:radius_cone}.
	This figure  shows the excess PDF of $\mu_1$ and $\nu$ for different satellite-to-central galaxy separations restricted to the first sample, for which trends are mutually exclusive, as seen in the top panels of Fig.~\ref{fig:green-orange}.
	The transition between the filamentary trend far from the central galaxy and the coplanar trend in its vicinity is striking, with a $50\%$ excess of satellites within the $37^\circ$ double cone around the filament axis ($30\%$ of satellites instead of $20\%$ for random) 
	at $R_{\rm gs}>2 \, R_{\rm vir}$ -- and no detectable coplanarity with the central at that distance -- that progressively decreases and turns to a $20\%$ excess outside of the corresponding $66^\circ$ bicone for $R_{\rm gs}<0.25 \, R_{\rm vir}$, 
	associated to a $\xi=70\%$ excess at $\mu_{1}=0$. In other words, $59\%$ of satellites are located outside the $66^\circ$ bicone revolving around the central minor axis ($40\%$ for random).

The investigation of satellite alignment with distance to the central galaxy carried out in this subsection shows that the {\it coplanar trend} is not a mere consequence of the {\it filamentary trend} as satellites transition from one trend to the other, 
	with  coplanarity  dominant in the vicinity of the central galaxy and  filamentary more prominent  in the outskirts of the halo.
	Hence, the dynamical bias introduced by the {\it filamentary trend} can reach an amplitude comparable to that of the {\it coplanar trend} for satellites within $R_{\rm vir}<R<2\, R_{\rm vir}$ from their central galaxy, an effect also seen 
	when the same analysis is performed on the full sample of centrals, regardless of orientation (see Appendix~\ref{section:separation-full}).
 
\begin{figure}
\center \includegraphics[width=0.7\columnwidth]{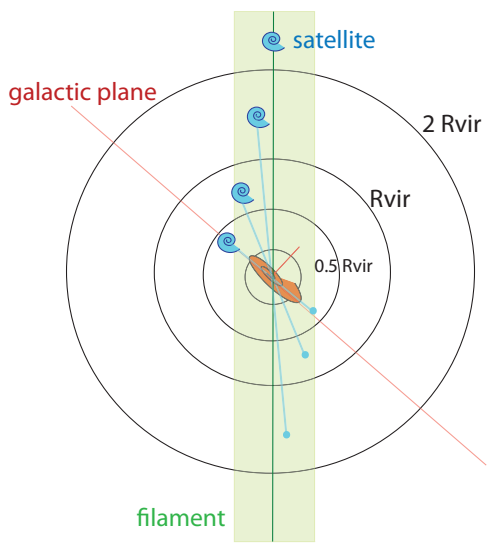}
  \caption{Sketch of the evolution of the alignment trends with distance to the center of the halo. In the outskirts of the halo, satellite galaxies are strongly aligned with the nearest filament. 
	Probing deeper into the halo this trend weakens as the alignment of satellites with the galactic plane strengthens.
  }
\label{fig:sketch2}
\end{figure}

An illustration of the satellite alignment evolution as a function of distance to the central  is shown in Fig.~\ref{fig:sketch2}. The transition between the coplanar and filamentary trends may represent a real source of angular bias in the distribution of satellites. In other words, inferring the anisotropy of the DM halo from the distribution of satellite galaxies assuming this distribution is unbiased may lead to significant errors. Effects of such errors will then strongly depend on the orientation of the central galaxy: whether its minor axis is aligned or not with the closest filament.

As a consequence, the transition between coplanar and filamentary trends for centrals with misaligned spin significantly impacts the statistics of the orientation of satellites around all centrals, as can be seen in Fig.~\ref{fig:radius_tot} in 
	Appendix~\ref{section:separation-full}, which displays the excess PDF of $\mu_1$ and $\nu$ for different satellite-to-central galaxy separations for the full sample irrespectively of the orientation of the central spin. Further analysis of this transition for 
	different central mass bins can be found in Appendix.~\ref{section:separation-full}. This Appendix highlights specific features that confirm the overall transition between filamentary infall and realignment of satellites through torquing deeper into the halo.

Note that these results suggest that the contradictory findings from observations since \cite{Holmberg69}'s  are mass and scale dependent, as \cite{Zaritsky97} suggested. 
  
\subsubsection{Interpretation}

The main results of this section suggest a dynamical scenario in which satellites flow along filaments and plunge into the halo where their orbits/spins are progressively deflected from alignment with the filament by gravitational torquing from the central galaxy, so as to 
	preferentially lie in the central galactic plane. In effect, their fate is reminiscent of that of the cold gas at higher redshift ($z>1$), which slithers as cold streams down to the core of central galaxies in formation 
	\citep{Pichon11,Codis12,Tillson15,danovichetal15}. High-redshift gas inflow in the frame of the galaxy is qualitatively double helix-like along its spin axis \citep{Pichon11}. Generated via the same winding/folding process as the protogalaxy, 
	it represents the dominant source of filamentary infall at redshift $z\simeq 2-3$ which feeds the galaxy with gas with well aligned angular momentum \citep{Pichon11,Stewart2013,danovichetal15}.

	Here, we argue that the distribution of satellites at $z<0.8$  traces that of the cold gas at $z>1$, directly correlated with the fact that satellites have progressively formed within these gas streams. This is not completely obvious a priori, as the gas, 
	unlike the satellites, can shock in the circumgalactic medium. In this picture, satellites initially aligned with the filaments in the vicinity of the halo end up corotating  in the central galactic plane, in agreement with \cite{danovichetal15} 
         who argue that central gravitational torques dominate even for the cold gas. To test this scenario, we  investigate possible kinematic signatures of such a trend in the next section.

\section{Kinematics of satellites}
\label{section:dynamics}

Let us now quantify  the tendency for satellites to corotate with their central galaxy, i.e. their tendency to align their orbital momentum (direction and polarity) with the intrinsic angular momentum of this central galaxy and  synchronize their circular velocity with its velocity.
\subsection{Corotation with the central galaxy}
\label{section:corot}

\begin{figure*}
\center \includegraphics[width=2\columnwidth]{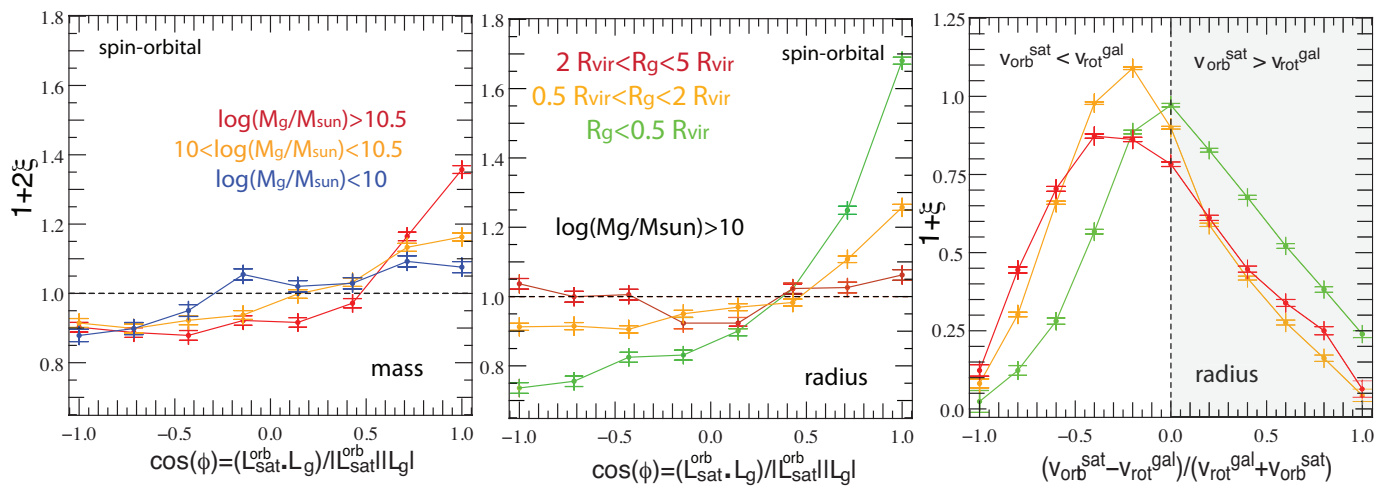}
 \caption{Twice the PDF of $ \cos \phi$, the cosine of the angle between the spin of the central galaxy and the total orbital momentum of its satellites for $0.3<z<0.8$ for different mass (left panel) and radius $R_{\rm gs}$ bins (middle panel). For massive central galaxies, the orbital 	
	angular momentum of satellites tends to align with the galactic spin of the central, especially  in its vicinity. Right panel: excess PDF of $( {v_{\rm orb}^{\rm sat}-v_{\rm rot}^{\rm gal}})/({v_{\rm orb}^{\rm sat}+v_{\rm rot}^{\rm gal}})$ for different radius bins, with $v_{\rm orb}^{\rm sat}$ 
	and $v_{\rm rot}^{\rm gal}$ the circular velocities of the satellites system and of the central galaxy respectively. Satellites increase their circular velocity and synchronize with their central as they reach the inner halo.}
\label{fig:corotation}
\end{figure*}

	To test the importance of the intrinsic torques of the central galaxy on its orbiting satellites, we first study their rotation around the central galaxy. Fig.~\ref{fig:corotation} shows the rescaled PDF, $1+2\xi$, of $ \cos \phi$, the cosine of the angle between the spin of the 
	central galaxy and the total orbital momentum of its satellites calculated in its rest-frame, for $0.3<z<0.8$. We study rotation evolution with mass in the left panel, but also rotation evolution as a function of distance to the central, $R_{\rm gs}$, in the middle panel. 

	Evolution as a function of mass and radius confirms the tendency of satellites to rotate on the galactic plane of their central with a circular velocity of the same sign as the central galaxy. This trend is observed for satellites of massive centrals
	($M_{\rm g} > 10^{10} \rm M_{\odot}$) and within $2 \, R_{\rm vir}$, and is all the more pronounced as the mass of the central increases and the distance $R_{\rm gs}$ decreases. 
	More massive central galaxies, and therefore host haloes, influence the orbital angular momentum of satellites more strongly: the excess probability at $\cos \phi=1$ is $3$ times higher for centrals with $M_{\rm g} > 10^{10.5} \rm M_{\odot}$ than for centrals 
	with $M_{\rm g} <10^{10} \rm M_{\odot}$. 

	Focusing on evolution as a function of distance to the central galaxy, we observe that $21.5\%$ of satellites within a $R_{\rm gs} < 0.5 \, R_{\rm vir}$ sphere around their central display an orbital angular momentum that remains within a $40^\circ$ double cone around its spin 
	and rotate in the same direction (to be contrasted to $12.5\%$ for a random distribution). In contrast, counter-rotation is more unlikely in the vicinity of the central galaxy, with only $8\%$ of the sample counter-rotating within a double cone of $40^\circ$ ($12.5\%$ for random).
	The relative orientation of the satellite's orbital angular momentum and the spin of the central galaxy is close to a random distribution outside the halo of the central galaxy, where satellites motions are governed by the filamentary flow.

	These results are consistent with a transfer of satellite orbital angular momentum to the intrinsic angular momentum of their host halo and central galaxy, through dynamical friction and gravitational torques.
	This exchange of angular momentum drives the evolution of the orbital angular momentum satellites, which end up corotating on the galactic plane, as they are dragged deeper into the halo. 
	This dynamical effect is what drives the {\it coplanar trend} in the central regions of the halo.
	 As such, it is a kinematic signature of the distinction between the coplanar trend at small distance and the filamentary trend at large distance.
	 	
To confirm this dynamical picture, we study the evolution of three quantities derived from definitions given in Section~\ref{section:kin}:
\begin{itemize}
\item The velocity contrast $( {v_{\rm orb}^{\rm sat}-v_{\rm rot}^{\rm gal}})/({v_{\rm orb}^{\rm sat}+v_{\rm rot}^{\rm gal}})$ with $v_{\rm orb}^{\rm sat}$ and $v_{\rm rot}^{\rm gal}$ corresponding to the angular orbital velocities of the system of satellites and to the rotational velocity of the central galaxy respectively. This quantity measures the tendency of satellites to synchronize their orbital velocity with the stellar material of the central galaxy.
\item  $L_{zc}/||\mathbf{L}_{\rm sat}^{\rm orb}||$,  with $L_{zc}$, the component of the orbital angular momentum of satellites aligned with the spin of the central galaxy and $||\mathbf{L}_{\rm sat}^{\rm orb}||$, the norm of the total orbital angular momentum. This quantity measures the tendency of systems of satellites to align their mean rotation plane with the central galactic plane.
\item $\sigma_{\rm plane}/||\mathbf{L}_{\rm sat}^{\rm orb}||$, with $\sigma_{\rm plane}$ the dispersion component. This quantity measures the dispersion of satellites' orbits around the mean orbit of the system for a given central, and, therefore, quantifies the tendency of satellites to be located in a thin rotation plane.
\end{itemize} 
The right panel of Fig.~\ref{fig:corotation} shows the PDF of $( {v_{\rm orb}^{\rm sat}-v_{\rm rot}^{\rm gal}})/({v_{\rm orb}^{\rm sat}+v_{\rm rot}^{\rm gal}})$ for different distance bins (with same color-coding as in the middle panel). The vertical dashed line separates the left (white) side of the panel where the satellite orbital velocity is lower than the central rotation velocity, and the right side (grey area) where satellites orbit faster. In the outer region of the halo, $R_{\rm gs}>0.5 \, R_{\rm vir}$, the average orbital velocity of satellites around their central galaxy is found to be lower than the rotation velocity of the central galaxy but $v_{\rm rot}^{\rm sat}$ increases in the inner part of the halo. Therefore, satellites increase their orbital velocity, synchronize with their central galaxy as they reach the inner part of the halo and achieve corotation. 

Two effects are competing: conservation of angular momentum tends to increase the amount of orbital velocity as satellites goes deeper in the halo, but the dynamical friction forces the orbital motions of satellites to synchronize with the rest of the matter in the halo, and this effect is stronger in the densest regions of the halo.

\begin{figure*}
\center \includegraphics[width=2\columnwidth]{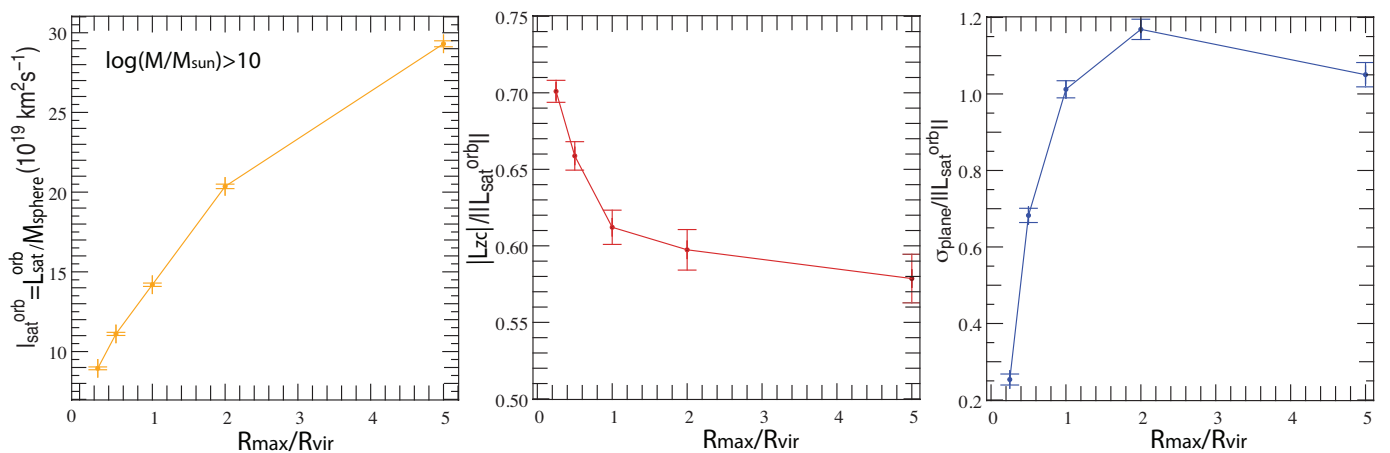} 
 \caption{{\it Left panel:} Average strength of the total specific orbital angular momentum $l_{\rm sat}^{\rm orb}$ of the system of satellites contained within  a sphere of radius $R_{\rm max}$  centred on their central galaxy. {\it Middle panel:} Evolution of the ratio between the 
	component of the satellite orbital momentum aligned with the spin of the central galaxy over the norm of the total satellite orbital angular momentum. {\it Right panel:} Evolution of the dispersion of orbital angular momentum $\sigma_{\rm plane}$ over the norm of the total 
	satellite orbital angular momentum for systems of satellites enclosed in spheres of increasing radius around the central galaxy. 
	The specific orbital angular momentum of satellites decreases for satellites closer to the central galaxy, and the relative importance 
	of the aligned angular momentum component increases for satellites closer to the central galaxy, while the relative amplitude of the dispersion between satellite orbits decreases.
 }
\label{fig:corotation4}
\end{figure*}

As expected from dissipation of angular momentum via dynamical friction in the halo, the strength of the average specific orbital angular momentum of systems of satellites $l_{\rm sat}^{\rm orb}=||\mathbf{L}_{\rm sat}^{\rm orb}||/M_{\rm sphere}$, with $M_{\rm sphere}$, 
	the total mass of satellites enclosed within a sphere of radius $R_{\rm max}$ from the central galaxy, (represented in the left panel in Fig.~\ref{fig:corotation4}) decreases for satellites closer to the central galaxy (by a factor 3 between satellites within $R_{\rm gs}< 5 \,R_{\rm vir}$ 
	and those within $R_{\rm gs}< 0.5 \, R_{\rm vir}$).
	Comparing between the ratios of $L_{zc}/||\mathbf{L}_{\rm sat}^{\rm orb}||$ (in the middle panel of Fig.~\ref{fig:corotation4}) and of $\sigma_{\rm plane}/||\mathbf{L}_{\rm sat}^{\rm orb}||$  (in the right panel of Fig.~\ref{fig:corotation4}), we see that the relative importance 
	of the aligned angular momentum component increases for satellites closer to the central galaxy (from 60\% to $70\%$), while the relative amplitude of the dispersion between satellite orbits drops from $120\%$ to $20\%$ between $2\, R_{\rm vir}$ and $0.25\, R_{\rm vir}$.
	This behaviour once again indicates that orbits of satellites progressively become coplanar due to torques. 
	On average, satellites lose orbital angular momentum as they are dragged deeper into the halo (decrease in $l_{\rm sat}^{\rm orb}$) but this trend is different along the different components of the orbital angular moment: the component aligned with the spin of the central galaxy is better preserved (even slightly increased) as satellites reach the inner parts of the halo than the unaligned components. 
	As a result, the dispersion in the distribution of satellites orbital planes as they 
	approach the central galaxy  shrinks.

\subsection{Satellite spin swings into the halo}
\label{section:satswings}

\begin{figure}
\center \includegraphics[width=0.85\columnwidth]{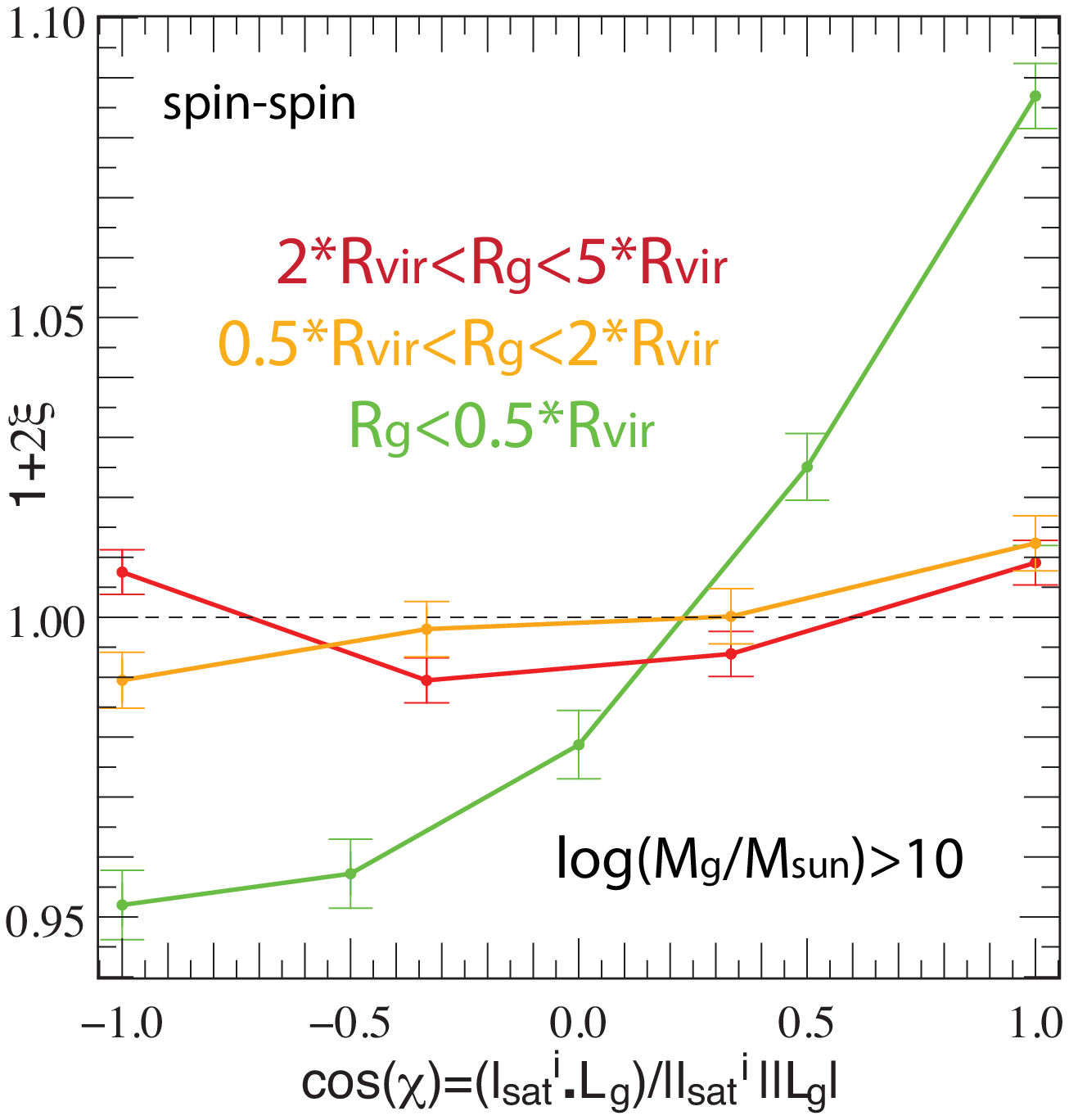} 
 \caption{Twice the PDF of $ \cos \chi$, the cosine of the angle between the spin of the central galaxy and the spin of the satellite for $0.3<z<0.8$ for different bins of distance to the central galaxy. Satellites align their intrinsic angular momentum to that of the central galaxies in 
	the inner part of the halo.}
\label{fig:corotation2l}
\end{figure}

As can be seen in Fig.~\ref{fig:corotation2l}, we also find that satellites not only align their orbital plane with the plane of the central galaxy, but also align their spin (intrinsic angular momentum) with that of the central galaxy as they reach the inner parts of the halo 
	\citep[see also][]{Aubert04}. In fact, cuts in mass and distance to the central galaxy lead to similar results as for $\cos \phi$, when applied to $\cos \chi$, the cosine of the angle between the central galaxy spin and the satellite spin, though the signal is  weaker and 
	rapidly decreasing with distance to the central. Nonetheless, within a $0.5 \, R_{\rm vir}$ sphere around the central, satellites have a $\xi=9\%$ excess probability to stay within a $37^\circ$ double cone around the spin of their central ($22\%$ of the satellites). This effect is 
	weaker than the previous trends but this excess was also found to reach $18\%$ for satellites within $0.25 \, R_{\rm vir}$, and even $20\%$ for the most massive central galaxies with $M_{\rm g} > 10^{11} \, \rm M_{\odot}$). Note that this particular measurement is at least 
	partially sensitive to grid-locking  (i.e. tendency of spins to align with the grid on which the gas fluxes are computed). Such effects are discussed in a companion paper, \cite{Chisari15}, which analyses the impact of grid-locking on intrinsic alignment measurements in more detail. 

\begin{figure}
\center \includegraphics[width=0.85\columnwidth]{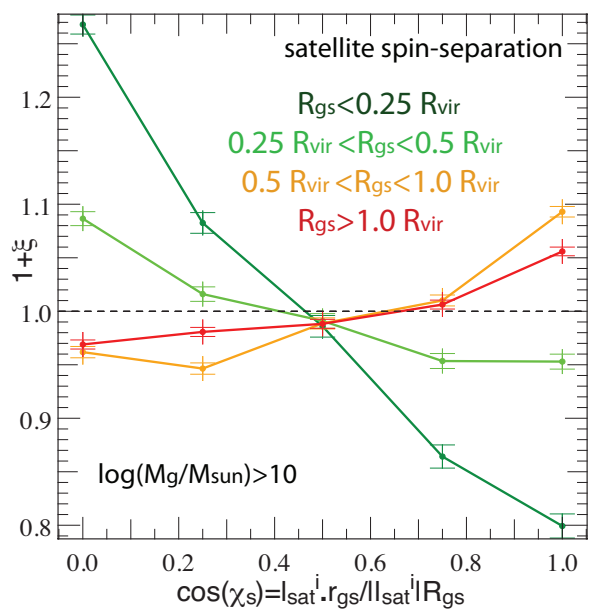} 
 \caption{PDF of $ \cos \chi_s$,  the cosine of the angle between the spin of the satellite and its position vector relative to the central galaxy as a function of distance to the central (radius bins are indicated on the figure).}
\label{fig:corotation2r}
\end{figure}

	Such satellite swings are confirmed by the measurement of the PDF of $ \cos \chi_s$,  the cosine of the angle between the spin of the satellite and its position vector relative to the central galaxy for different separation bins (see Fig.~\ref{fig:corotation2r}). We find that outer 
	satellites have a spin preferentially aligned with their position vector. This is consistent with low-mass infalling galaxies having their spin aligned with the filament they are flowing from~\citep{duboisetal14}. Conversely inner satellites have a spin more likely to be perpendicular 
	to their position vector, hence a rotation plane aligned with it. Note that this result holds for the full sample of central galaxies with $M_{\rm g}>10^{10}\,M_{\odot}$, irrespective of the fact that it is dominated by centrals with a galactic plane well aligned or moderately misaligned 
	with the nearest filament. Satellites progressively swing their rotation plane to align with that of the central as they reach the inner parts of the halo. 

	These results  statistically confirm  the importance of torquing from the massive central as a driving mechanism  for satellite alignment.
	Such torques are in agreement with theoretical predictions derived from linear response theory by \cite{Colpi98-2}.
	This author interpreted the orbital decay as triggered not by the surrounding halo, but by the central galaxy (stellar material) itself on its external satellites,  via near resonance energy and angular momentum transfers. This mechanism noticeably leads  to a circularization of orbits, 
	and an alignment between the major axis of the satellite with $\mathbf{r}_{\rm gs}$, a result also found for dark haloes in N-body simulations by \cite{Aubert04} and \cite{Faltenbecher08}. 
	We show below in Section.~\ref{section:colorsat} that the evolution of the age of satellites as traced by colours also lends support to this interpretation of our results.

	Several observational works have looked for signs of satellite alignment from their shapes. \cite{Schneider13} find a weak signal of radial alignment of satellites in observations from the `Galaxy And Mass Assembly' survey. This $3\sigma$ detection is obtained when projected galaxy shapes are modelled using S\'ersic profiles. The significance decreases to $2\sigma$ when up-weighting the inner parts of galaxies (as for weak lensing shape measurements). \cite{Sifon15} and \cite{Chisari14} recently looked at projected ellipticity alignments around stacked clusters at low redshift ($0.05<z<0.5$) and did not find any significant alignment. This may suggest that satellite alignments in most massive clusters are mostly damped by non-linear evolution or projection effects, at least in the local Universe. In contrast \citet{Singh14} found a significant one-halo component to the alignment signal of luminous red galaxies in the SDSS survey. The discrepancy between these works could be caused by a luminosity dependence of the alignment signal, along with the use of different radial weights for recovering galaxy ellipticities. In comparison, our shape estimation is un-weighted, and thus it is presumably more sensitive to tidal features in the outskirts of galaxies. Also, we have relied up to now on three-dimensional galaxy shapes, while projection is known to damp the alignment signal, as we will see in the next section.

\section{Mock observations}
\label{section:mock}

	In this section, we provide further insight into how the alignment signal can be traced in observations and how our trends compare to what is found in existing surveys. 
 
 \subsection{Colors in Horizon-AGN}
\label{section:colagn}

	In most observational studies, the mass of galaxies and satellites is traced by their rest-frame colours. It is therefore of interest to recover the previously described variation of the satellite alignment trends with the mass of the central galaxy, using central galaxy rest-frame colours.
\begin{figure*}
\center \includegraphics[width=1.9\columnwidth]{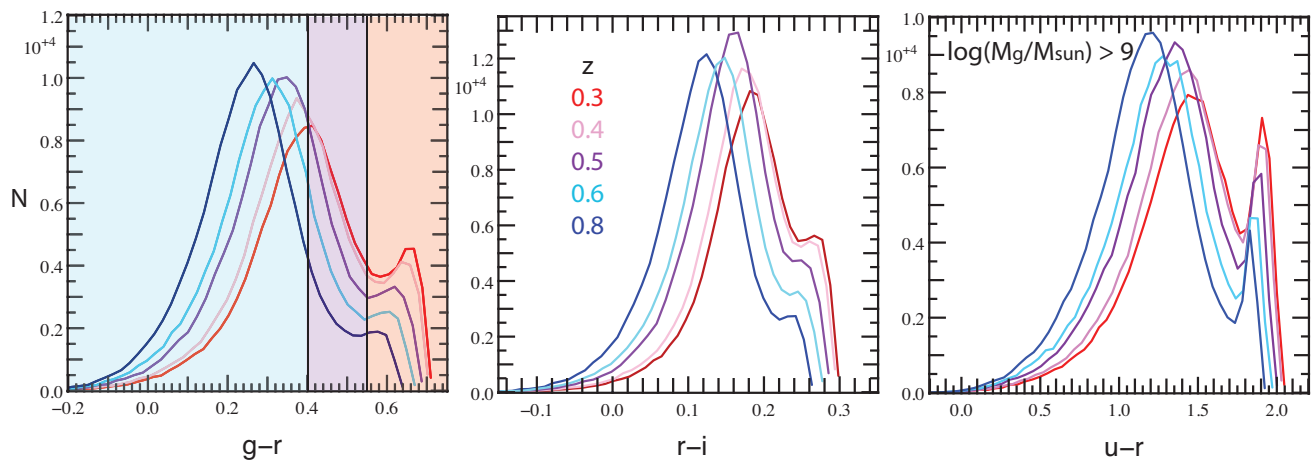}
 \caption{Distribution of rest frame colours in Horizon-AGN. Coloured areas indicate the colour bins used throughout the analysis in this section of the paper.}
\label{fig:bimodality}
\end{figure*}
%
	To trace in the observations  the dynamical scenario previously described, we need to make a distinction between young outer satellites and old(er), quenched, inner satellites. To estimate the age of satellites we rely on 
	their rest-frame colours computed from  AB magnitudes in SDSS filters.
	Fig.~\ref{fig:bimodality} shows this distribution for three rest-frame colours, $g-r$, $r-i$ and $u-r$  in Horizon-AGN for all galaxies with $M_{\rm g} > 10^{9} \, \rm M_{\odot}$ at different redshifts. Note that these colours do not take into account dust extinction.
	Although the blue cloud-red sequence bimodality is not as clear in the simulation as it is in observed colour-mass diagrams, the simulated galaxy population reveals a noticeable increase in the fraction of red galaxies as redshift decreases (more galaxies 
	are passively evolving).
	The cuts used in the remainder of the paper to split the galaxy population into blue (star-forming), intermediate (green valley) and red (quenched) galaxies are represented in Fig.~\ref{fig:bimodality} (left panel) and correspond to $g-r<0.4$, $0.4<g-r<0.55$ and $g-r>0.55$. 
	They are defined in this way as a compromise between isolating the two peaks present in the colour distribution and selecting large enough galaxy samples. We use the same colour cuts to split central and satellite galaxies.

\subsection{Alignment trends with color selection}
\label{section:selcol}

\subsubsection{Results in three-dimensions}

\begin{figure*}
\center \includegraphics[width=1.7\columnwidth]{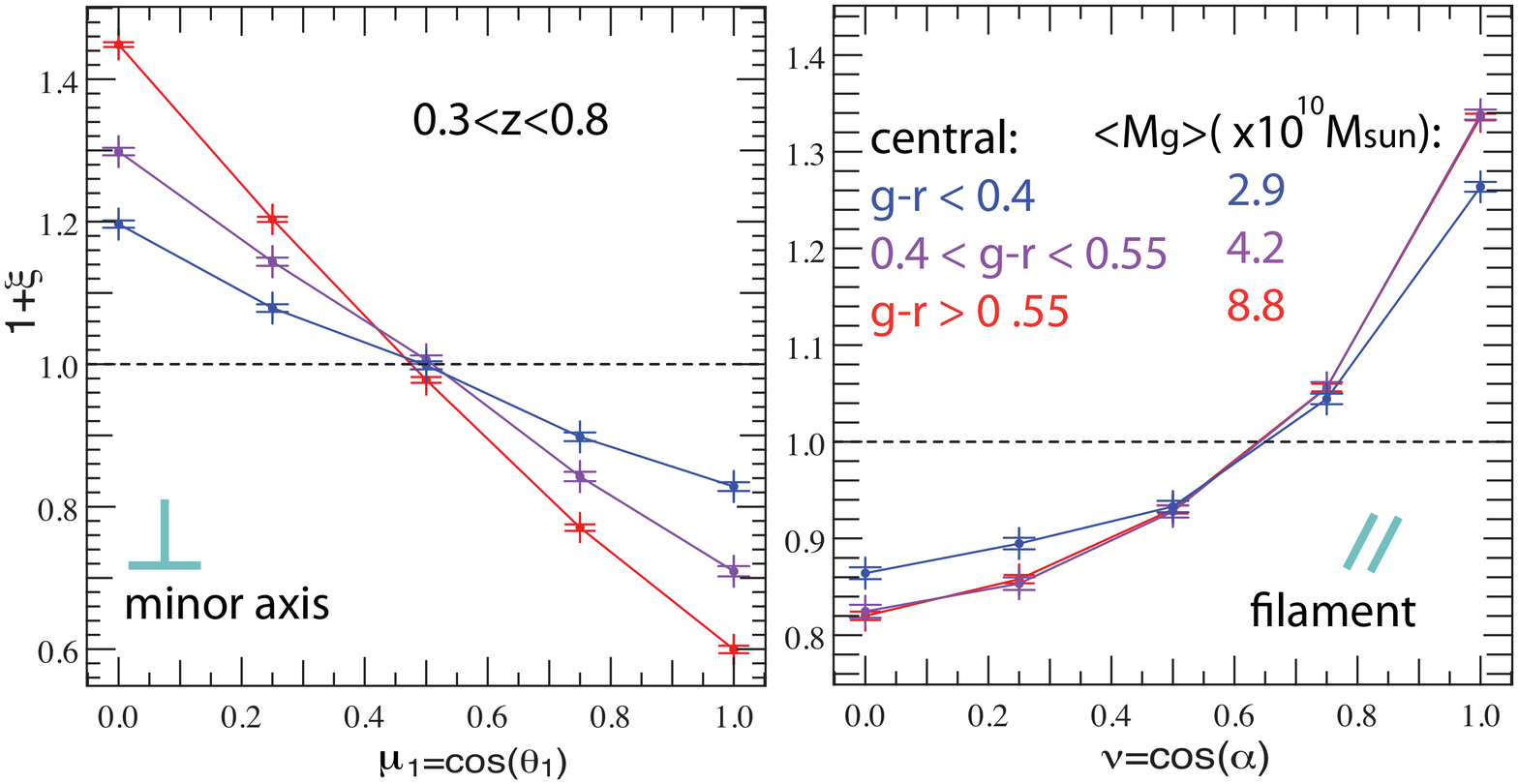}
 \caption{{\it Left panel}: PDF of $ \mu_{1}=\cos \theta_{1}$, the angle between the minor axis of the central galaxy and the vector separating it from its satellites, at $0.3<z<0.8$ and for different colour bins. {\it Right panel}: PDF of $ \nu=\cos \alpha$, the angle between the satellite
	 separation vector and the direction of the nearest filament. The average central galaxy mass in each colour bin is also indicated. {\it Right panel}: PDF of $ \mu=\cos \theta$, the angle between the spin of the central galaxy and the satellite separation vector. For massive red 
	central galaxies, the satellites  tend to be distributed on the galactic plane. The filamentary trend is also recovered although degraded by projection for the bluest central colour range.}
\label{fig:colcut}
\end{figure*}

	 Fig.~\ref{fig:colcut} shows two plots very similar to the mass-dependent plots presented in the previous sections (e.g. Fig.~\ref{fig:mass0}), albeit where mass bins are replaced by $g-r$ color bins for the central galaxy. 
	As red central galaxies are older and more massive their blue counterparts, we expect to observe an increase in the {\it coplanar trend} as  $g-r$ increases. This is indeed the case, as can be seen in the first panel of the figure, which displays the PDF of 
	$ \mu_{1}=\cos \theta_{1}$ for three different central galaxy colour bins, and for all satellites within a sphere of radius $5\,R_{gs}$ centred on these centrals. Red centrals with $g-r > 0.55$ tend to have their satellites aligned in their galactic plane, with $54\%$ of satellites 
	outside the $66^\circ$  bi-cone revolving around the central minor axis, an amount which falls down to $46\%$ for blue centrals with $g-r < 0.4$ (by comparison the value expected for a uniform distribution is $40\%$). 
	The {\it filamentary trend} is also observed in the right panel, with an excess of probability similar to that when the trend is broken down in mass instead of colour. The fact that blue centrals are more likely to be 
	young galaxies with a spin parallel to the filament explains why blue centrals are subject to a slight decrease in the {\it filamentary trend} compared to their red counterparts: they are more likely to be found in a situation where both trends compete.

	As a conclusion, color selection proves as efficient as mass selection to identify and quantify both trends, which is consistent with a steady evolution of the average mass in each color bin for all galaxies with $M_{\rm g}>10^{10} \, \rm M_{\odot}$: {\it red} galaxies have an 
	average mass of $8.8\times 10^{10} \, \rm M_{\odot}$, while it falls down to $4.2\times 10^{10} \, \rm M_{\odot}$ for the {\it intermediate} bin and $2.9\times 10^{10} \, \rm M_{\odot}$ for the {\it blue} galaxies. 

	Additionally, the PDF 
	of $ \mu=\cos \theta$, (central spin-separation angle) can be found in Appendix. While using the spin rather than the minor axis does not change our results qualitatively, the amplitude of 
	the spin signal is significantly lower. The discrepancy between those two signals is highly dependent on the shape of the central galaxy, which can induce significant misalignments between the minor axis and the spin, as suggested in the next section. 
%

\begin{figure*}
\center \includegraphics[width=1.7\columnwidth]{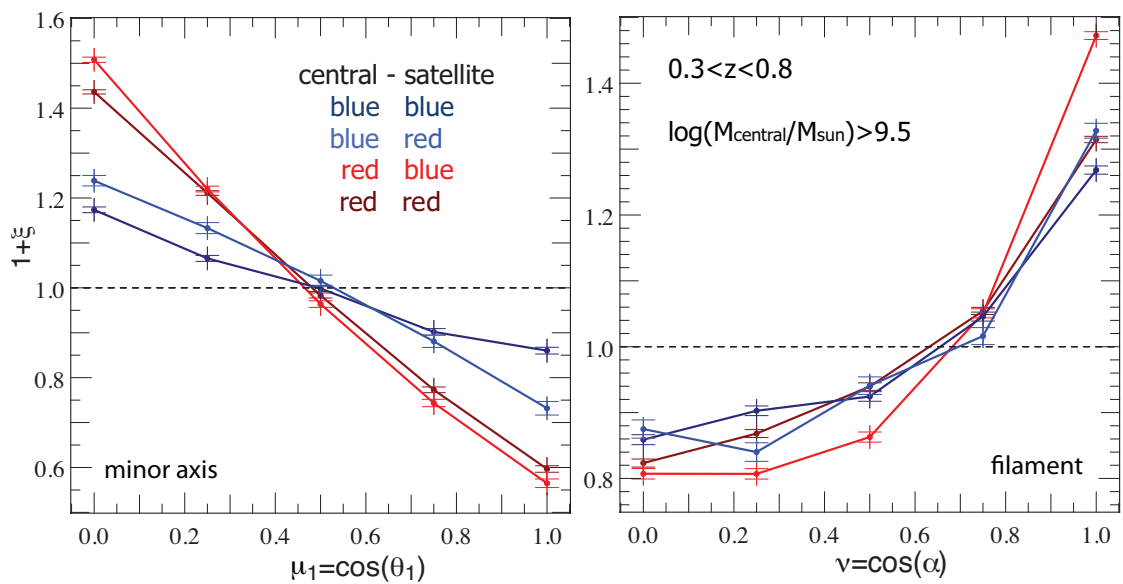}
 \caption{Same as Fig~\ref{fig:colcut} but splitting both the central galaxies {\em and} the satellites in different colour bins. Blue galaxies are identified as objects with $g-r < 0.4$ and red galaxies as objects with $g-r > 0.55$.}
\label{fig:colcol}
\end{figure*}

Finally, in Fig.~\ref{fig:colcol} we compare our results to observations by \cite{Yang06} and simulations by \cite{Dong14}, who also performed a detailed analysis of the alignments of galaxies as a function of the colours of both satellites and centrals. We find that:
\begin{itemize}
\item The {\it coplanar trend} is stronger for red centrals, especially when they have bluer satellites (although this distinction is quite minor). We attribute this to more efficient torques exerted by massive central galaxies, as the mass ratio $m_{\rm sat}^{i} / M_{g}$ 
	is smaller on average for blue satellites. It is important to remember that trends are more likely to reinforce in this case, as red centrals are often dominated by mergers in their stellar mass budget as opposed to in situ star formation~\citep{Oser10, Dubois2013}, 
	and hence have a higher chance of maintaining a spin orthogonal to their filament. Therefore, the distance to the filament is not crucial in this case, as satellites fall directly from the filament into the galactic plane.
\item Blue centrals, younger and less massive, are more likely to have a spin parallel to their filament -- which induces a competition between the {\it filamentary} and the {\it coplanar} trends -- and less likely to efficiently torque satellites to line them up in their rotation plane. 
	Expectedly, the {\it coplanar} signal is weaker than that for their red counterparts. Consequently, the signal is then slightly stronger for red satellites of blue centrals which are more evolved and closer on average to their central than blue satellites. 
\end{itemize}

Those results are consistent with those of \cite{Yang06} who report a significant alignment of satellites along the major axis of their (projected
 central. Although these authors found a red-red signal higher than the red-blue one, their study focused 
	on $R_{\rm gs}< R_{\rm vir}$ which potentially left aside an important number of blue satellites in alignment with the filament (recall that we use all satellites up to $5\, R_{\rm vir}$ separation). However, their general colour and mass trends for the central galaxy are in good agreement with our results, as detailed in the next section.

\begin{figure*}
\center \includegraphics[width=1.8\columnwidth]{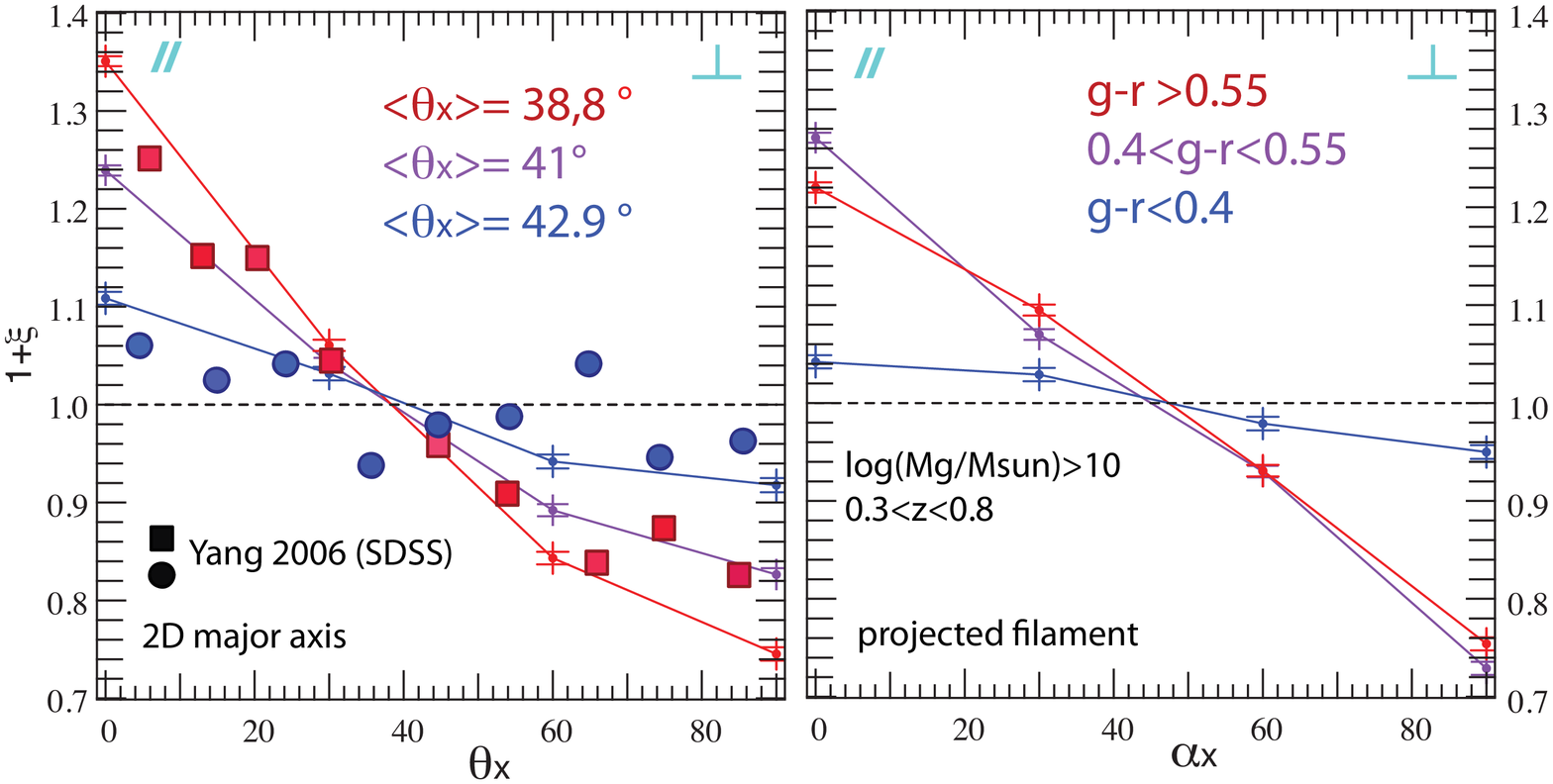}
 \caption{PDF of $\theta_{x}$ and $\alpha_{x}$, the angles between the $x$-projected major axis of, respectively, the central galaxy (left panel), and direction of the filament (right panel) and the $x$-projected $\mathbf{r_{\rm gs}}$, at $0.3<z<0.8$ for different color bins. For massive red central galaxies, the satellites tend to be distributed on the galactic plane. The projected signal is qualitatively comparable to results in 3D. Observational results from \citet{Yang06} are overplotted for red centrals (red squares) and blue centrals (blue dots). The agreement is good.}
\label{fig:col2d}
\end{figure*}

\subsubsection{Alignments in projection}

Fig.~\ref{fig:col2d} shows alignments projected along the $x$-axis of the grid. The left panel shows the PDF of $\theta_{x}$ the angle between the major axis of the projected central galaxy and the projected $\mathbf{r_{\rm gs}}$, at $0.3<z<0.8$ for different colour bins. 
	The right panel displays the PDF of $\alpha_{x}$, the angle between the projected direction of the filament and the projected $\mathbf{r_{\rm gs}}$. Results are in good agreement with the observed signal found in the SDSS by \cite{Yang06} (overplotted as blue dots and red squares on Fig.~\ref{fig:col2d}), although alignment trends seem to
	 be slightly stronger in our case, when accounting for the fact that our mass range is biased towards smaller masses.  Average values for $\theta_{x}$ in each colour bin are given in the left panel and confirm the steady evolution of the trend with $g-r$. This increase is sharper than that found in \cite{Yang06}, however we believe the results remain consistent given that we do not model dust extinction, which likely impacts our estimation of galaxy colours.
%
%
{ \center
\begin{table}
\begin{tabular}{|ll|c|c|c|}
\hline
\multicolumn{2}{|c|}{} & \multicolumn{3}{c|}{${<\theta_{\rm X}>}$ ($^{\circ}$)} \\
 \hline
{central} & { satellite} &  ${R_{\rm gs} < 5\, R_{\rm vir}}$ & ${R_{\rm gs}< R_{\rm vir}}$ & { Yang 06}  \\
 \hline
red & red & 38.9 & 40.0 & 40.8 \\
\hline
red & blue & 38.4 & 39.3 & 42.9 \\
\hline
blue & red & 41.9 & 41.9 & 44.8 \\
\hline
blue & blue & 43.1 & 42.1 & 44.2 \\
\hline
\end{tabular}
\caption{Mean values (in degrees) for $\theta_{x}$ in different color bins for both satellites and central galaxies, and within two different radius from their central galaxy. Values are systematically compared to observed values by \citet{Yang06}.}
\label{fig:table}
\end{table}
}
%

Extinction notwithstanding, it is interesting to notice that the projected estimation follows very closely the 3D results, although it leads to a systematic slight underestimate of the alignment trends.  

Repeating the measurement of Fig.~\ref{fig:colcol} in projection along the $x$-axis of the grid, we also recover similar alignments that result in a mean angle variation detailed in Table ~\ref{fig:table}. This mean angle is presented for satellites within two different 
	maximum separation from their central galaxy: $R_{\rm vir}$ and $5\, R_{\rm vir}$, and systematically compared to values found in \cite{Yang06}. Colour variations are qualitatively preserved and  suggest that the transition between the {\it coplanar} and {\it filamentary} trend with separation can be recovered in projection, although 
	the effects are weak and an analysis of the central spin's orientation might be necessary.

\subsection{Evolution of satellites within the halo}
\label{section:colorsat}

Let us now focus on the satellite dynamical transition from a filamentary to a coplanar distribution as they plunge into their host halo.

\begin{figure}
\centering{ \includegraphics[width=0.85\columnwidth]{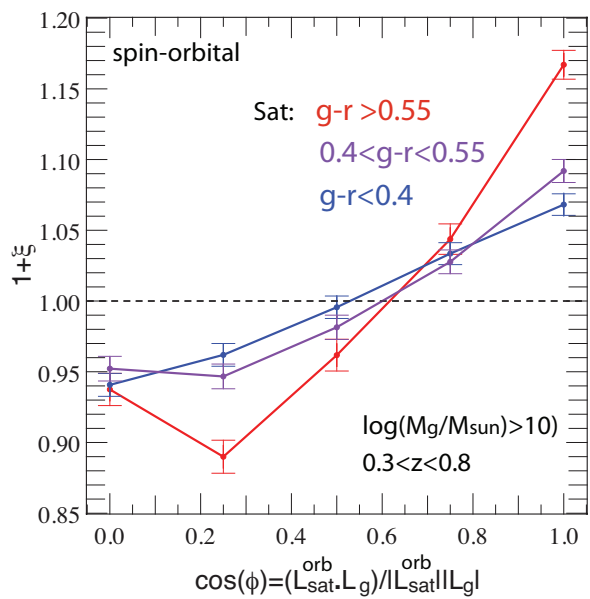}}
 \caption{PDF of $|\cos \phi|$, the cosine of the angle between the spin of the central galaxy and the total orbital momentum of its satellites (measured in the rest frame of their central) for $0.3<z<0.8$ and for different $g-r$ satellite colour bins. 
	For red quenched satellites, the orbital momentum tends to be aligned with the central galactic spin. The effect is much weaker for blue star forming satellites, consistent with their preferential orientation along the filament.
 }
\label{fig:sat_co}
\end{figure}

\begin{figure}
\centering{ \includegraphics[width=0.9\columnwidth]{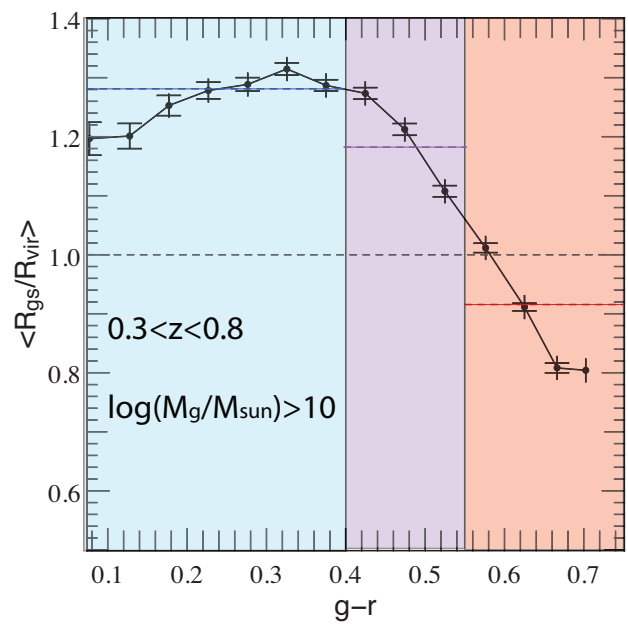}}
 \caption{Evolution of the mean (rescaled) central-satellite separation as a function of the satellite $g-r$ colour (for all satellites within $R_{\rm gs}<5\,R_{\rm vir}$). Our colour bins are overplotted as coloured areas and the average separation in each bin is indicated by a dashed line. 
 }
\label{fig:sat_co_r}
\end{figure}

	Fig.~\ref{fig:sat_co} shows the PDF of $\cos \phi$, the angle between the spin of the central galaxy and the total orbital angular momentum of its satellites for $0.3<z<0.8$ and for different satellite $g-r$ colour bins. We see that red satellites, on average, have an orbital plane better 
	aligned with the central galactic plane than blue satellites ($24\%$ of the sample within the $37^\circ$ double cone around the spin of the central for $g-r>0.55$, and $21.5\%$ for $g-r<0.4$). 
	Moreover, Fig.~\ref{fig:sat_co_r} presents the average distance of satellites to the central as a function of their colour for satellites within $5\, R_{\rm vir}$: red galaxies are closer to the central ($\simeq 1.3\, R_{\rm vir}$) than blue galaxies ($\simeq0.9 \,R_{\rm vir}$). 
	Therefore, red satellites are more clustered around the central galaxies than blue satellites, as an effect of ram-pressure stripping and strangulation which gradually remove gas from satellites and prevent further accretion onto them as they evolve in the hot 
	pressurised atmosphere of the host halo.

\begin{figure}
\center \includegraphics[width=0.85\columnwidth]{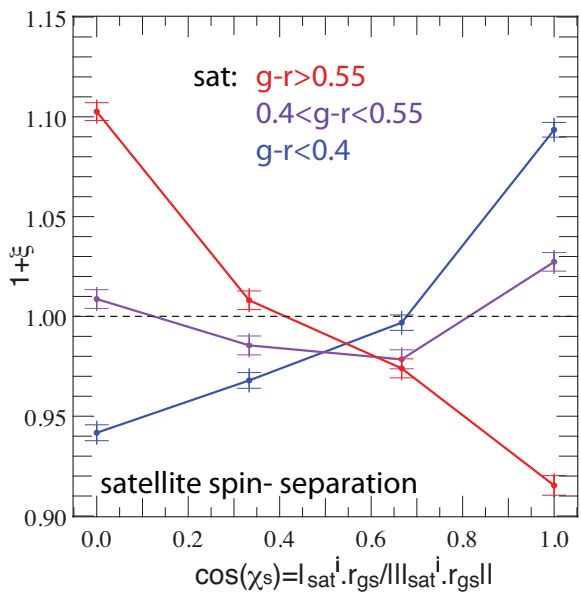} 
 \caption{PDF of $ \cos \chi_s$,  the cosine of the angle between satellite spin and position vector for different satellites in  $g-r$ colour bins.
Blue (outer) satellites have a spin preferentially
	 aligned with their position vector, while red (inner) satellites have a rotation 
	plane aligned with it.
 }
\label{fig:corotation2col}
\end{figure}

	Hence, using colours, we also recover the trend that satellite orbits achieve better coplanarity with the central galaxy as they get closer to it and get more and more deprived of star forming gas.
%
%
%
To confirm  the satellite rotation plane swings,  we plot in Fig.~\ref{fig:corotation2col} the PDF of $ \cos \chi_s$,  the cosine of the angle between the spin of the satellite and its position vector for different colour bins. We find that blue (outer) satellites have a spin preferentially
	 aligned with their position vector, in accordance with their spins being aligned with the filament they are flowing from, while red (inner) satellites have a spin more likely to be perpendicular to their position vector, hence a rotation 
	plane aligned with it.

\begin{figure*}
\center \includegraphics[width=2.0\columnwidth]{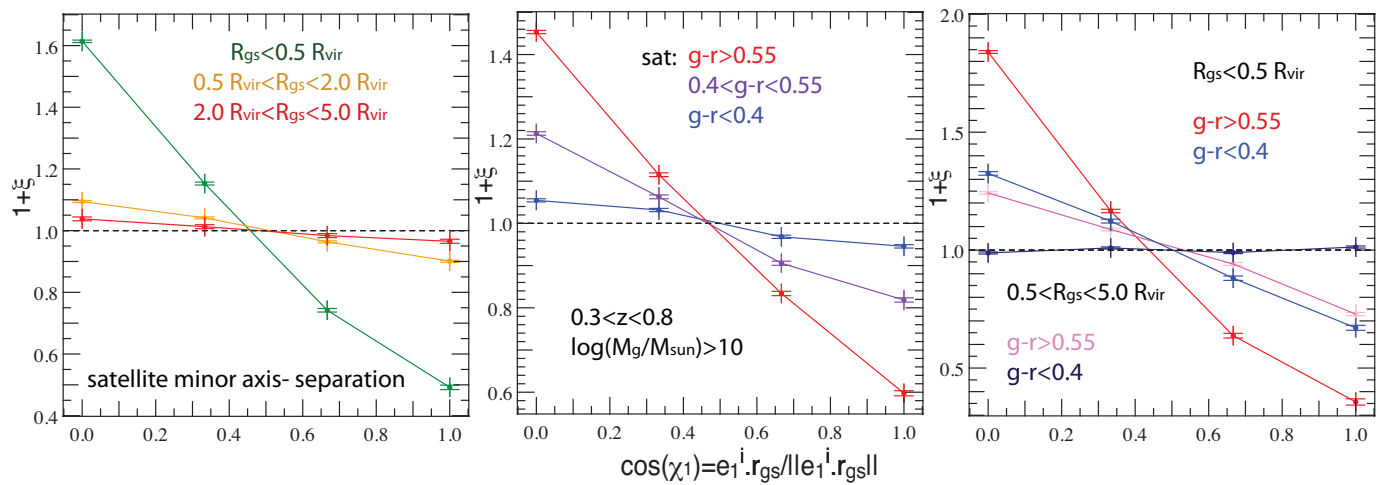} 
 \caption{PDF of $ \cos \chi_1$, the cosine of the angle between the minor axis of the satellite and its position vector for $0.3<z<0.8$ for different radius (left panel), satellite colour (middle panel) and mixed (right panel). Satellites align their minor 
	axis with that of the central galaxy as they reach the inner regions of the halo.}
\label{fig:corotation3}
\end{figure*}

	In Fig.~\ref{fig:corotation3}, we consider the orientation of the minor axis rather than that of the spin of the satellite for different distance and colour bins. This static geometrical parameter is more sensitive to limited resolution (recall that the minimum number of stellar particles is lower for satellites than for centrals), stripping and friction than the orientation of the satellite spin 
	and we do not detect a flip as clear as the one found for the spin, but the evolution is globally similar and the tendency to display a minor axis orthogonal to the galactic plane for redder satellites in the inner parts of the halo is strengthened. It confirms the dynamical mechanism 
	that bends the rotation plane of satellites to align it with their orbital plane. As this latter progressively aligns itself with the central galactic plane, the rotation plane of satellites also ends up aligned.
	
\subsection{Shape effects}
\label{section:ellipse}

\begin{figure}
\center \includegraphics[width=0.9\columnwidth]{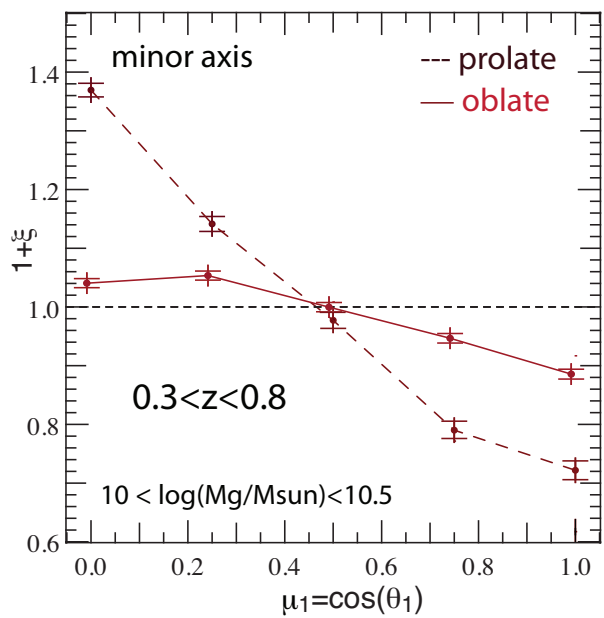}
 \caption{PDF of $ \mu_{1}=\cos \theta_{1}$ similar to that in Fig.~\ref{fig:mass0} for oblate (solid line) and prolate (dashed line) central galaxies, restricted to the intermediate mass range. Tendency to lie orthogonal to the minor axis is stronger for prolate structures, as in this case the minor axis is most often orthogonal to the filament direction and the galactic plane is poorly defined.}
\label{fig:shape}
\end{figure}

	A full understanding of the amplitude of the alignment signal requires to take into account its dependence on the shape of the central galaxy, as trends display different features for oblate (``disky") and prolate (``spheroidal") structures. 
	Fig.~\ref{fig:shape} shows the evolution of the PDF of $ \mu_{1}=\cos \theta_{1}$ (minor axis/separation angle) for oblate and prolate central galaxies in the intermediate mass range for which the deviation between both samples is maximal. 
	The satellite orthogonality to the minor axis is more pronounced for prolate structures. However, one should not deduce that it corresponds to a better alignment on the galactic plane, as such a plane for prolate structures is poorly defined. For prolate galaxies, the 
	rotation plane is more likely to be orthogonal to the major axis, hence not coinciding with the plane of maximal elongation.  Prolate structures show a significant amount of misalignment between their spin and minor axis, with more than $30 \%$ 
	displaying a spin aligned with the major axis.

%
The evolution of the PDF of $ \mu_{1}=\cos \theta_{1}$ and $\nu=\cos\alpha$  for satellites of central galaxies with a spin aligned to the filament (not represented here) confirmed that the alignment of satellites  along the minor axis of their central galaxy and along the filament cannot be straightforwardly deduced from the orientation of the spin for such prolate structures. In this case, only the {\it filamentary} trend is clearly detected, while other alignments are strongly dependent on the proxy used to define the galactic plane (shape or rotation plane) and how it correlates to the filament direction.

\begin{figure}
\center \includegraphics[width=0.8\columnwidth]{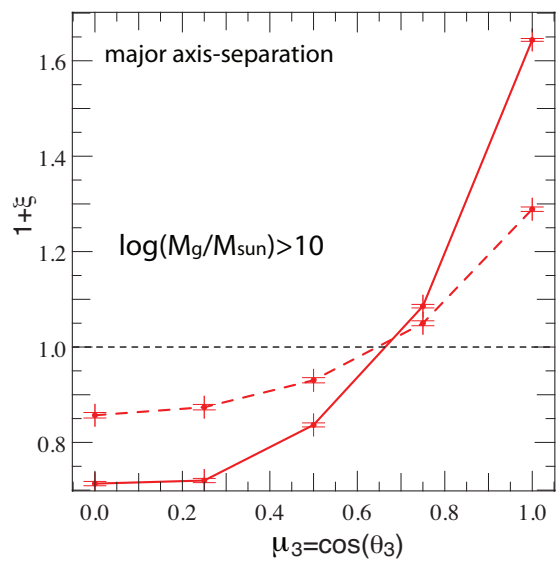}
 \caption{PDF of $ \mu_{3}=\cos \theta_{3}$ the cosine of the angle between $\mathbf{r_{\rm gs}}$ and the major axis of the central galaxy for oblate (dashed line) and prolate (solid line) centrals. Prolate centrals have their satellites aligned along their major axis, which is consistent with 
	the fact that this axis is more often aligned with the spin in this case. Oblate centrals display a certain degree of satellite alignment too, which is consistent with the distribution of satellites tracing the underlying triaxiality of their host halo.}
\label{fig:major}
\end{figure}

Finally, focusing on the correlations with the central shape, Fig.~\ref{fig:major} shows the PDF of $ \mu_{3}=\cos \theta_{3}$ the cosine of the angle between $\mathbf{r_{\rm gs}}$ and the major axis of the central galaxy for oblate (dashed line) and prolate (solid line) centrals. 
	While oblate centrals display a certain degree of satellite alignment along their major axis, prolate centrals have their satellites more strongly aligned with their major axis. This is consistent with a distribution of satellites tracing the underlying triaxiality of their host halo
	but is also consistent with the fact that the major axis is more often aligned with the spin in the prolate case, since prolate galaxies are very likely to be dispersion-dominated systems (a.k.a. ellipticals).
%

\section{Discussion}
\label{section:discussion}

In this paper, we have compared the alignment trends identified in \hagn with observational constraints obtained mainly from the SDSS. We have found good agreement with the work of \cite{Yang06}, who confirmed the tendency of satellite galaxies to align on the galactic plane of their central galaxy and found this trend to be mass and color dependent (for both central galaxies and satellites), and \cite{Zaritsky97} who found it to be also separation (distance) dependent. Our results also confirm the satellite galaxies' tendency to align with their host's closest filament. This trend is found to be independent of the orientation of the central galactic plane at large separations, as established in recent observations by \cite{Tempel15} \citep[see also][]{Paz11}. Closer kinematic analysis reveals a transition for satellites from filamentary infall at large separations to orbit alignment in the central galactic plane in the vicinity of their central, when such a plane is misaligned with the closest filament. This dynamical behavior is very similar to that of the cold gas in high-redshift zoom simulations \citep{Tillson15,danovichetal15} and suggests a tight connection between the dynamics of cold gas accretion in the early universe and the alignment trends of satellites in clusters in the Local Universe.

The development of models to quantify such trends is crucial for upcoming imaging surveys to achieve their goals of constraining the equation of state of dark energy and modifications to General Relativity, such as {\it Euclid}\footnote{\url{http://sci.esa.int/euclid}}~\citep{Laureijs11}, the Large Synoptic Survey Telescope\footnote{\url{http://www.lsst.org}}\citep{Ivezic08} and WFIRST\footnote{\url{http://wfirst.gsfc.nasa.gov/}}~\citep{Spergel13}. Hydrodynamical cosmological simulations are a promising tool to quantify intrinsic alignments in the nonlinear regime and provide estimates of contamination to future surveys (\citealt{Tenneti14a,Tenneti14b,Chisari15}). The results of simulations can be used to inform the parameters of an intrinsic alignment ``halo model'' \citep{Schneider10}, or nonlinear models that rely on perturbation theory power spectra \citep{Blazek15}. 

However, our work shows that alignments of galaxies on small scales are the result of a complex dynamical interplay between the central galaxy, the satellites and the surrounding filamentary structure. In short, alignment trends depend on the evolutionary stage of a galaxy (as probed by e.g. its colour) as well as on the orientation of the central with respect to the nearest filament. Morever, our work suggests that alignment trends exist for ``blue discs'', and their significance may be increased for the less evolved galaxies at the higher redshifts that will be typically probed by weak lensing surveys. While this alignment signal tends to be suppressed in projection, its potential contamination to lensing remains to be explored. We have also shown that alignments can transition between two regimes as satellites move from the surrounding filament into the 
	gravitational zone of influence of the central galaxy. The {\it filamentary trend} implies a tangential aligment of discs around centrals, resulting in a tangential shear signal that adds to the galaxy-lensing in projection. The {\it coplanar trend} represents a net radial orientation of satellites 
	and their central, suggesting that galaxy-lensing could be suppressed on the small scales in projection. Both trends would contribute to a cosmic shear measurement through correlation of intrinsic shapes and weak lensing \citep[the `GI' term][]{Hirata04}. 
	
	A companion paper,
	 \citet{Chisari15}, studies projected correlations of intrinsic shapes in the language of lensing measurements, including the case of a two-halo term (the correlation between centrals).
These authors  studied alignments on scales $>100 \, h^{-1}\, \rm kpc$, finding an overall tendency for discs to align their rotation planes tangentially around spheroidals, which we now re-interpret as consequence of the {\it filamentary trend}. 
They also found a tendency for spheroidals to be elongated towards other spheroidals, which can be a consequence of this type of galaxies being elongated following the direction of the filament, which connects them to other spheroidals. In comparison, \citet{Chen15} also found evidence for alignment of the major axis of massive galaxies in the direction of the nearest filament (also, nearest group) in the MassiveBlackII smoothed-particle-hydrodynamics cosmological simulation. However, no evidence has been found for tangential alignment of discs in that simulation nor in the moving-mesh Illustris simulation \citep{Tenneti15b}; a discrepancy that will require further investigation. Finally, \citet{Chisari15} suggested that the spin and the shape of a spheroidal are poorly correlated, a conclusion that we validated for prolate galaxies in Section \ref{section:ellipse}.

\section{Conclusion and prospects}
\label{section:conclusion}

	We investigated the alignment of satellites galaxies in the redshift range $0.3<z<0.8$ . We found alignments on the galactic plane consistent with previous investigations (both observational and numerical), 
	but we also unraveled an interesting dynamical interplay between this known alignment and the tendency of satellites to be aligned with the filament from which they flow into the halo. Noticeably, we find that, although most massive central galaxies display some coplanarity 
	among their satellites, this effect is enhanced when the galactic plane lies parallel to the direction of the filament. Conversely, it can be weakened or even canceled when the galactic plane is misaligned with its filament. Satellites are then found to lie preferentially along the 
	filament far from the central galaxy, and on the galactic plane in its vicinity.

\begin{figure}
\center \includegraphics[width=0.85\columnwidth]{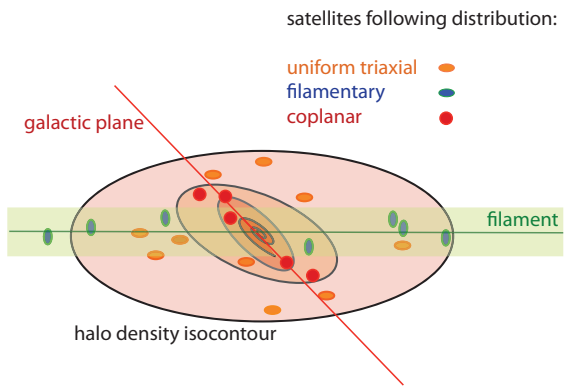}
 \caption{Summary of the orientation trends that drive the fate of satellites in their host halo. Alignment trends are distance dependent: {\it filamentary} trend dominates in the outskirt of the halo (for bluer satellites on average), while {\it coplanar trend} takes over in the vicinity of the central (redder satellites on average). These trends may bias the standard expectation that the satellites distribution traces the overall shape of the diffuse dark matter halo (orange ellipses). More probably, specific populations of satellites (selected in mass, color and separation) trace specific parts of the halo and the embedding cosmic web.}
\label{fig:recap}
\end{figure}

The main results of this work are sketched  in Fig.~\ref{fig:recap}, where we illustrate the fact that the distribution of satellites in their host halo and around their central galaxy arises from the superposition of at least two different effects, represented in addition to the common expectation that they should trace the tri-axiality of the halo:
\begin{itemize}
\item[i)] the polar flow from the filament, which mostly affects young blue satellites in the outskirt of the halo.
\item[ii)] the dissipation in the halo and torques from the central galaxy, which bend older inner redder satellite orbits close to coplanarity with the galactic plane. 
\end{itemize}
A natural question that arise from these trends is whether such a satellite distribution is still in good correspondence with the triaxiality of the host halo, and how it might affect estimations of the halo triaxiality on specific scales.

%
%
	The analysis of the radial  evolution of those  processes in Horizon-AGN strongly suggests that:
	\begin{enumerate}
	\item The leading effect in the orientation of satellites is the tendency to align with the nearest filament. While stronger for satellites in the outskirts of the halo, this tendency decreases as the satellites are dragged deeper into the halo -- where they exchange angular momentum  
	via gravitational torques from the central galaxy -- but not to the point where it becomes negligible, unless a strong misalignment ($>45^{\circ}$) exists beforehand between the central galaxy minor axis and the filament direction.
	\item A secondary effect that becomes dominant in the inner parts of the halo is the tendency of satellites to align with the central galactic plane. This effect can either compete or strengthen the alignment with the filament, depending on the orientation of the plane.
	 As expected, the signal is stronger for red massive centrals which are already more likely to have a spin orthogonal to the filament, and therefore a galactic plane aligned with the filament. The two effects reinforce in this case.  On the contrary, low mass blue centrals 
	with a minor axis aligned with their filament have satellites predominantly aligned with their filament and so orthogonal to the galactic plane.
	\item The alignment of  satellites in both the filament and the galactic plane is consistently stronger for red central galaxies, as these correspond to massive centrals. The dependence of this {\it coplanar} trend on the $g-r$ colour of satellites is kinematically consistent with a 
	dynamical scenario in which young (blue) satellites flowing from the filament progressively bend their orbits towards the central galactic plane (under its tidal influence). As they reach the inner parts of the halo and they progressively get deprived of more of their gas and stars 
	through tidal (and ram pressure) stripping, becoming redder in the process. We predict this trend is likely to be observable.
	\item Around $40\%$ of massive centrals with $M_{\rm g}>10^{10}\, \rm  M_{\odot}$ display significant deviations from the spin-filament orthogonality ($>30^{\circ}$) and are therefore subject to competing alignment trends between the two processes, with the alignment 
	along the filament predominant for blue satellites in the outskirts of the halo, and coplanarity with the central galaxy taking over for older red satellites in the inner regions of the halo. 
	\item The tendency for systems of satellites to align on the galactic plane is accompanied with a tendency to align and synchronize their orbital momentum with the angular momentum of the central galaxy. We also find hints that satellites align their intrinsic angular 
	momentum with that of their central  as they reach the inner regions of their host halo.
	\end{enumerate}

In conclusion, this investigation has shown that the distribution of satellites in host halos is dynamically biased first by the filamentary anisotropic flow from which they originate and second by the gravitational torques of the central galaxy which bend them close to the galactic plane. This evolution closely resembles that of the cold gas, even though the satellite population does not shock in the circumgalactic medium. 
Given that such flows were identified at high redshift, it is remarkable that it has a stellar counterpart at low redshift in the possibly observable colour variation of the satellite distribution. 
The anisotropy of the infall and realignment within the virial radius as characterized in this paper could have an impact on building up thick discs, turning discs into ellipticals and producing either fast or slow rotators~\citep{atlas3D}.

\begin{acknowledgements}
This work was granted access to the HPC resources of CINES (Jade and Occigen) under the allocation 2013047012, c2014047012 and c2015047012 made by GENCI.
This research is part of the Spin(e) (ANR-13-BS05-0005, \url{http://cosmicorigin.org}) and Horizon-UK projects. 
Let us thank D.~Munro for freely distributing his {\sc \small  Yorick} programming language and opengl interface (available at \url{http://yorick.sourceforge.net/}). 
We  warmly thank S.~Rouberol for running  the {\tt Horizon} cluster on which the simulation was  post-processed.  Part of the analysis was also performed using the DiRAC facility jointly funded by BIS and STFC.
The research of JD is supported by Adrian Beecroft, The Oxford Martin School and STFC. 
CP thanks Churchill college for hospitality while this work was finalized.
\end{acknowledgements}
\vspace{-0.5cm}

\bibliographystyle{aa}
\bibliography{author}

\appendix


\section{Low-mass centrals}
\label{section:low-mass-centrals}

We stated in Section.~\ref{section:mass-align} that the overall anisotropic distribution of satellites around central galaxies is, for a good part, inherited from  filamentary infall and from the mass-dependent orientation of central galaxy planes relative to the nearby filament. 
	However, this naively implies that satellites of low-mass centrals ($M_{\rm g}< 10^{10}\,\rm M_{\odot}$) should preferentially be orthogonal to the central plane, while the distribution of alignments we find for this mass range is closer to random, see Fig.~\ref{fig:mass0}. 
	In this Appendix, we detail the reasons for this apparent discrepancy.
%
\begin{figure}
 \includegraphics[width=0.85\columnwidth]{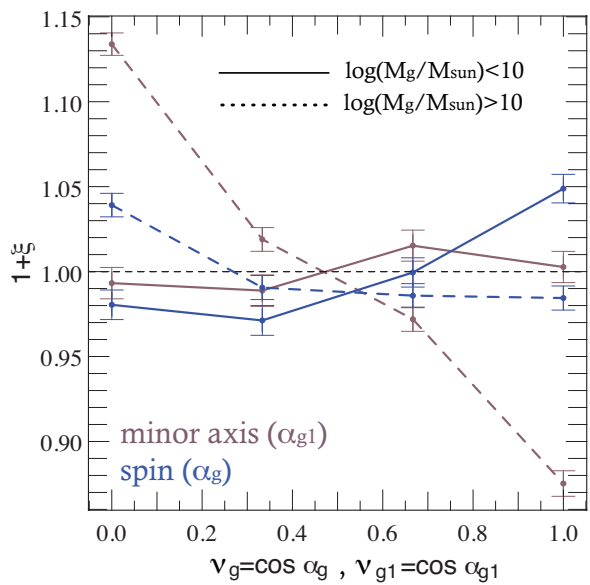}
  \caption{PDF of  $\nu_{g1}=\cos \alpha_{g1}$, the cosine of the angle between the minor axis of the central galaxy and the direction of the nearest filament (brown lines) and $\nu_{\rm g}=\cos \alpha_{\rm g}$, the angle between the spin of the central galaxy and the direction of 
	the nearest filament  (blue lines). Distributions are plotted for central galaxies in two different mass ranges: $10^{9}<M_{\rm g}<10^{10} \, \rm M_{\odot}$ (solid lines) and $M_{\rm g}>10^{10}\, \rm M_{\odot}$ (blue lines). The transition for the galactic plane orientation is 
	recovered, but the signal is stronger using the spin for low-mass centrals. The opposite is true for high mass centrals: the minor axis estimator yields  an increased amplitude in the excess probability $\xi$.}
\label{fig:central_spin}
\end{figure}
In Section.~\ref{section:mass-align}, Fig.~\ref{fig:mass0} we opted for a comparison with the direction of the minor axis of the central galaxy to quantify alignment on the galactic plane. This is a well justified method for high-mass galaxies, sufficiently  well resolved to neglect 
	errors due to a lack of convergence   of the shape estimator with the number of particles, as it gives a stronger and smoother signal:  many of them  have little rotation relative to their dispersion support, often displaying strong variations between the inner and outer shells of 
	the galaxy as a result of their on-going merging activity. Moreover, most of these massive centrals  have satellites of comparatively smaller mass which have little impact on the shape of the central prior to merger.

	However, this statistics is more questionable  for less resolved low-mass centrals as such systems of  centrals with satellites are often pairs (or triplets) of interacting galaxies ($85\%$ of the sample, $97\%$ including systems with two satellites). As we define a threshold around
	 $10^{8.5}\, \rm M_{\odot}$ for structure detection and a threshold $M_{\rm g}=10^{9}\, \rm M_{\odot}$ for centrals, low-mass centrals have by construction fewer satellites. Consequently, these pairs have mass ratios close to one (pre-major merger systems).   Therefore, galaxies 
	in such pairs strongly impact each other's shape through tidal interactions, which results in stronger and faster variations of their central minor axis direction than what is expected for a reference sample of galaxies with $M_{\rm g}< 10^{10}\,\rm M_{\odot}$  without satellites. Comparatively, the spin direction, which somehow probes better the inner rounder parts of galaxies that do not weight much in the calculation of the inertia tensor, and must also remain consistent with the conservation of the total angular momentum of the pair proves more robust.

While previous studies of the Horizon-AGN simulation from \cite{duboisetal14} and \cite{Welker14}  were performed at $1<z<3$, we focus on a  lower redshift range: $0.3<z<0.8$. As a result, the tendency of massive galaxies to display a spin/minor axis perpendicular to their nearest
	 filament is expected to increase with cosmic time due to the cumulative effect of mergers. In a similar vein, the parallel orientation trend of young, low mass galaxies is expected to decrease at lower redshift as a result of the disappearance of the cold and vorticity rich gas inflows
	 that feed them. 
	To illustrate these effects, Fig.~\ref{fig:central_spin} displays the PDF of $\nu_{g}=\cos\alpha_{g}$ (blue lines), with $\alpha_{g}$ the angle between the spin of the central and the direction of the nearest filament, and $\nu_{g1}=\cos\alpha_{g1}$ (brown lines), with $\alpha_{g1}$ 
	the angle between the minor axis of the central and the direction of the nearest filament, for two mass ranges of central galaxies: $10^{9}<M_{\rm g}< 10^{10}\,\rm M_{\odot}$ (solid lines) and $M_{\rm g}>10^{10}\,\rm M_{\odot}$ (dashed lines). 
	The solid lines on the figure lend support to the claim that galactic planes of low-mass centrals are preferentially oriented orthogonally to the nearest filament,
	but that this alignment is better traced with the spin-filament angle than with the minor axis-filament angle (for which the signal is strongly damped and compatible with a random distribution).
	Conversely, when tracing the (parallel) orientation of the galactic plane of massive centrals 
	(dashed lines) with respect to the filament minor axis yields a stronger signal than spin. However, we emphasize that the signal for low-mass galaxies remains weak using either the spin or the minor axis to compute 
	the plane orientation, with a maximum amplitude $\xi=5\%$ for the excess probability.
.  
%
\begin{figure*}
 \includegraphics[width=1.8\columnwidth]{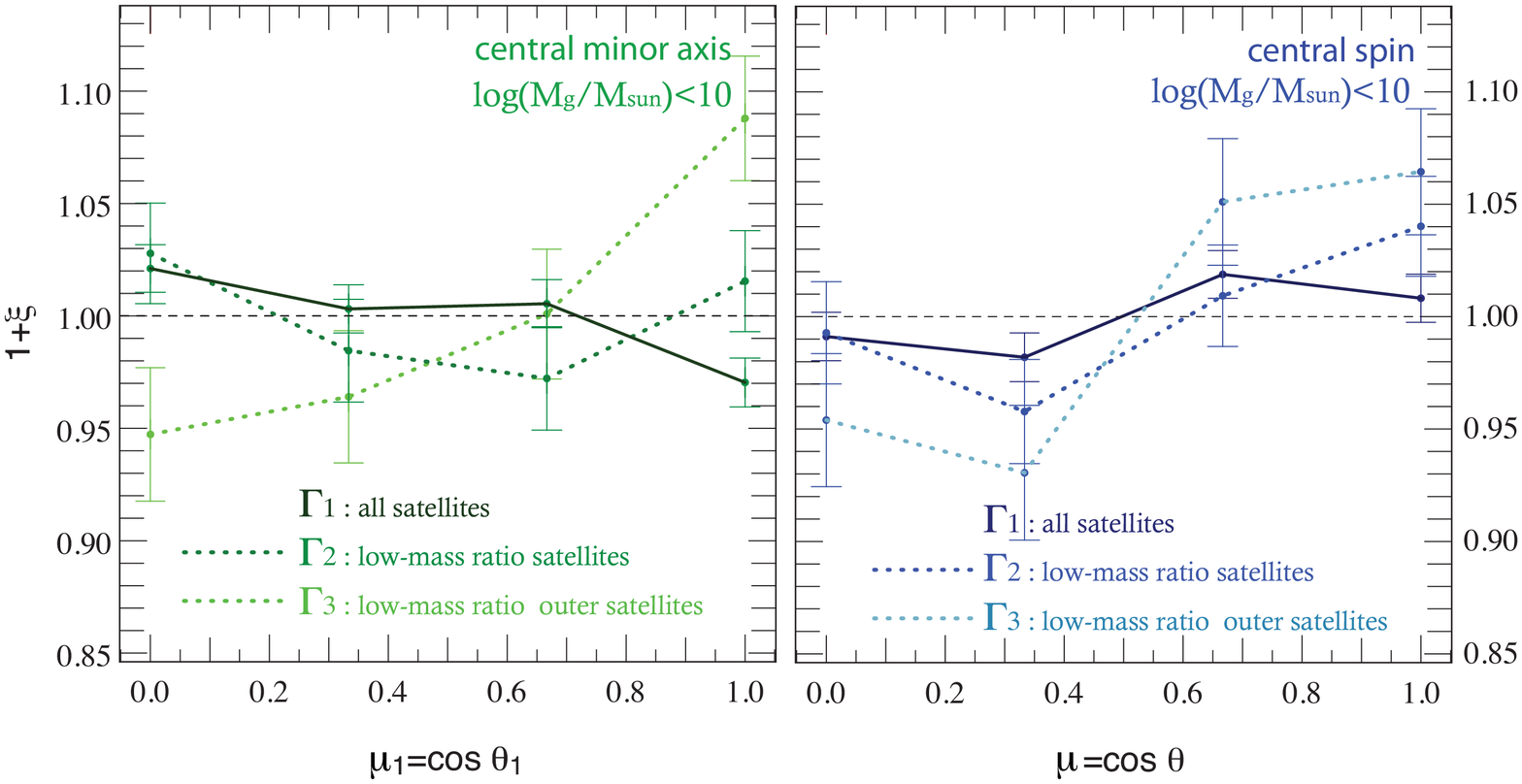}
  \caption{{\it Left panel:} PDF of $ \mu_{1}=\cos \theta_{1}$, the angle between the minor axis of the central galaxy and the satellite separation vector $\mathbf{r_{gs}}$. {\it Right panel:} PDF of $ \mu=\cos \theta$, the angle between the spin of the central galaxy and the separation vector. Distributions are plotted for three different subsets of low-mass centrals with varying types of satellites (green and blue shades): $\mathbf{\Gamma_{1}}$: $10^{9} <M_{\rm g}<10^{10}\, \rm M_{\odot}$, $\mathbf{\Gamma_{2}}$: $5\times10^{9}<M_{\rm g}<10^{10}\, \rm M_{\odot}$ and $M_{\rm sat}< 2 \times10^{9}\, \rm  M_{\odot}$,  $\mathbf{\Gamma_{3}}$: $5\times 10^{9}<M_{\rm g}<10^{10}\, \rm M_{\odot}$, $M_{\rm sat}< 2 \times 10^{9}\, \rm M_{\odot}$ and $R_{\rm gs}>R_{\rm vir}$. 
   Satellites of low-mass centrals tend to align with their central's spin/minor axis, consistently with the fact that they tend to align with the nearest filament. However, the signal is limited in amplitude and tends to vanish when traced with the minor axis for centrals with massive satellites and/or with satellites in their close vicinity. In this mass range for the central, the satellite is often of comparable mass and both galaxies equally impact the orientation and shape of one another prior to merger.
  }
\label{fig:lowm_central}
\end{figure*}
%

Fig.~\ref{fig:lowm_central} shows the PDF of $ \mu_{1}=\cos \theta_{1}$ (left panel, green lines) and of $ \mu=\cos \theta$ (right panel, blue lines) for the satellites of three different samples of centrals:  $\mathbf{\Gamma_{1}}$, which includes all the satellites around centrals in 
	the lower mass range: $10^{9}<M_{\rm g}<10^{10}\, \rm M_{\odot}$, $\mathbf{\Gamma_{2}}$ which focuses on centrals with low mass ratio satellites ($M_{\rm sat}/M_{\rm g}<0.4$): $5\times10^{9} <M_{\rm g}<10^{10}\, \rm M_{\odot}$ irrespective of their distance to the central and 
	$M_{\rm sat}< 2 \times10^{9}\, \rm M_{\odot}$, and $\mathbf{\Gamma_{3}}$ which focuses on centrals with low mass ratio, outer satellites only: $5\times 10^{9}<M_{\rm g}<10^{10}\, \rm M_{\odot}$, $M_{\rm sat}< 2 \times 10^{9}\, \rm M_{\odot}$ and $R_{\rm gs}>R_{\rm vir}$. 	
	Focusing on sample $\mathbf{\Gamma_{1}}$,  Fig.~\ref{fig:lowm_central} confirms the lack of a clear detection for satellites to be oriented orthogonally to their central galactic plane. Although a weak signal is detected when computing the spin-separation angle,  using  the 
	minor axis-separation angle yields an opposite signal of similar amplitude, i.e. a weak tendency for satellites to lie on the galactic plane of such centrals. However, when focusing on $\mathbf{\Gamma_{2}}$ and $\mathbf{\Gamma_{3}}$, i.e. centrals with no massive satellites and 
	centrals with neither massive nor close satellites, the perpendicular orientation of satellites is progressively recovered for both $\theta$ (for $\mathbf{\Gamma_{2}}$ and $\mathbf{\Gamma_{3}}$)  and $\theta_{1}$ (for $\mathbf{\Gamma_{3}}$).  This result supports the idea that 
	the damping of the low-mass central trend results from their greater sensitivity to interaction with satellites, amplified by their excess probability of having satellites of similar mass.
\begin{figure}
 \includegraphics[width=0.85\columnwidth]{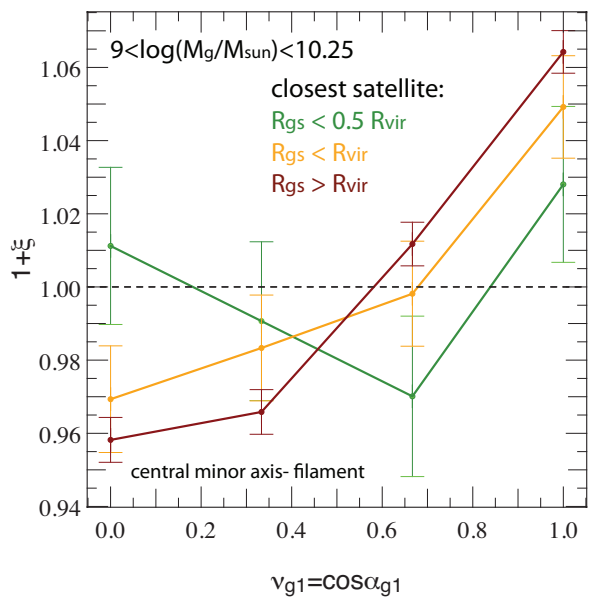}
  \caption{PDF of  $\nu_{g1}=\cos \alpha_{g1}$, the cosine of the angle between the minor axis of the central galaxy and the direction of the nearest filament, for three subsets of low-mass centrals ($10^{9}<M_{\rm g}<10^{10.25}\, \rm M_{\odot}$, upper threshold corresponds to the lower limit for the galactic plane orientation transition mass as estimated in \citealt{duboisetal14}) depending on the distance of their closest satellite: $R_{\rm gs}<0.5\, R_{\rm vir}$, $0.5\, R_{\rm vir}<R_{\rm gs}<R_{\rm vir}$ and $R_{\rm gs}> R_{\rm vir}$. The presence of a close satellite impacts the orientation of the central spin and damps its tendency to align to the nearest filament. The closer the satellite, the stronger the effect.}
\label{fig:with_sat}
\end{figure}

This effect is perhaps best emphasized when focusing on the PDF of $\nu_{g1}=\cos\alpha_{g1}$,  the cosine of the angle between the minor axis of the central galaxy and the direction of its nearest filament, for different subsets of centrals depending on whether or not their halo 
	hosts a close satellite. The results of  this analysis are presented in Fig.~\ref{fig:with_sat} which displays this PDF for centrals with $10^{9}<M_{\rm g}<10^{10.25}\, \rm M_{\odot}$.
	This threshold corresponds to the lower limit for the transition mass where galaxy spins switch between preferentially aligned and preferentially perpendicular to the closest filament, as estimated in \cite{duboisetal14}. Such a choice of threshold allows us to preserve a decent 
	statistics for each subsample. These centrals are further divided in three subsets: those whose 
	closest satellite is found within $0.5\, R_{\rm vir}$ (in green), those whose closest satellite is found between $0.5\, R_{\rm vir}$ and $R_{\rm vir}$ (in yellow), and those whose closest satellite is more distant than $R_{\rm vir}$ (in dark red). One clearly sees that the presence 
	of a satellite in the vicinity of a low-mass central clearly impacts the orientation of its spin, damping its tendency to align with the nearest filament. This effect is increased for closer satellites.

	This explains why the alignment of satellites orthogonally to the galactic plane of low-mass centrals is not measured in Fig.~\ref{fig:mass0}, even though they do follow the {\it filamentary} trend.

\section{Distance to the central galaxy}
\label{section:separation-full}

	Let us investigate here the effect of alignment change as a function of distance to the central for the full sample of galaxies, i.e. without specific mass cuts.
\subsection{Full sample}

	Quantifying the relative influence of the filament and the joint effect of angular momentum dissipation via dynamical friction in the halo and central galaxy torques (which also alter the inner halo shape) is best achieved by noticing that these processes 
	operate on different radial scales 
	\citep[as shown by][for cold flows]{danovichetal15}. We presented results for a sub-sample of centrals in Fig.~\ref{fig:radius_cone}.
	We now measure the distribution of satellite galaxies around their central as a function of distance for the full sample of central-satellites pair, with no further consideration on the spin orientation of the central galaxy.

\begin{figure*}
\center \includegraphics[width=1.5\columnwidth]{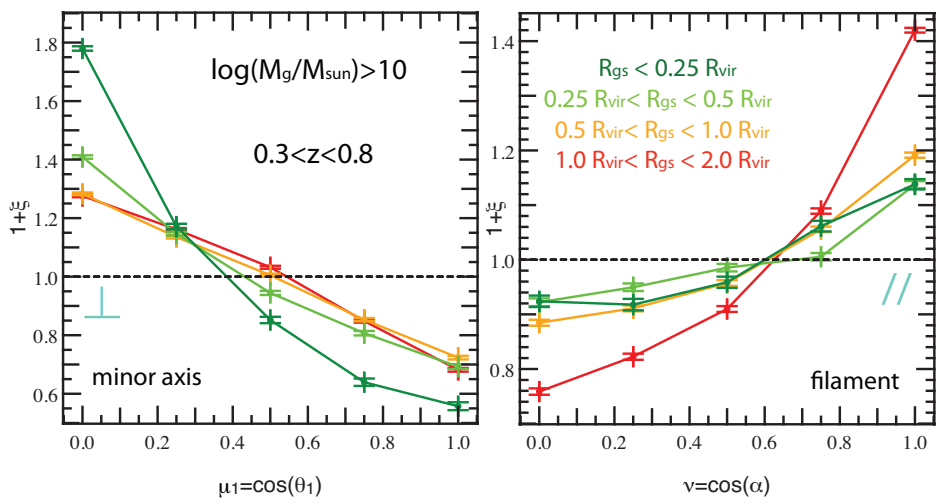}
  \caption{Same as Fig.~\ref{fig:mass0} where samples are binned in distance, $R_{\rm gs}$,  from satellite to central.
	Satellites close to the central galaxy tend to be distributed on the galactic plane and exhibit only marginal alignment with the filament, while satellites in the outskirts of the host halo of the central are strongly aligned with the filament, while their coplanarity with the central 
	galaxy is weakened. Results are stacked for $0.3<z<0.8$.
}
\label{fig:radius_tot}
\end{figure*}

	Fig.~\ref{fig:radius_tot} shows the excess PDF of $\mu_{1}=\cos \theta_{1}$ (central galaxy minor axis) and $\nu=\cos \alpha$ (filament axis) for different bins of $R_{\rm gs}$, the distance from the satellite to the central galaxy. 
	We see that, as the distance increases, the alignment with the galactic plane weakens progressively while the alignment with the filament is strengthened.
	For satellites in the vicinity of their central, within a sphere of radius $R_{\rm gs}=0.25 \, R_{\rm vir}$, the {\it coplanar trend} is highly dominant with $\xi=80\%$ at $\mu_1=0$, and with $60\%$ of the satellites outside a $66^\circ$  double cone around the minor axis.
	($40\%$ for random), while the {\it filamentary trend} is reduced to a $12\%$ excess within a $37^\circ$ double cone around the filament axis, which hosts $23\%$ of the satellites ($20\%$ for random).
	In contrast, for satellites in the outskirts of the halo with  $R_{\rm gs}>R_{\rm vir}$  the {\it filamentary trend}  dominates with this same excess, soaring to $41\%$ at $\nu=1$ and hosting $28\%$ of the satellites in a $37^\circ$ double cone around the filament axis 
	($20\%$ for random). The amount of satellites outside the $66^\circ$ double cone around the minor axis falls down to $48\%$ that case ($40\%$ for random).

	Thus, in this ``full-sample" analysis we recover the evolution with distance characteristic of a transition between both alignment trends for satellites around centrals whose spin is aligned to the nearest filament.
	However, in the present case, contrary to the results discussed in the main body of the paper, centrals both with aligned and misaligned spin are analysed together, which leads to an overall persistence of the 
	filamentary trend at any distance from the central galaxy. 
	
\subsection{Evolution for different mass bins}

\begin{figure*}
\center \includegraphics[width=1.4\columnwidth]{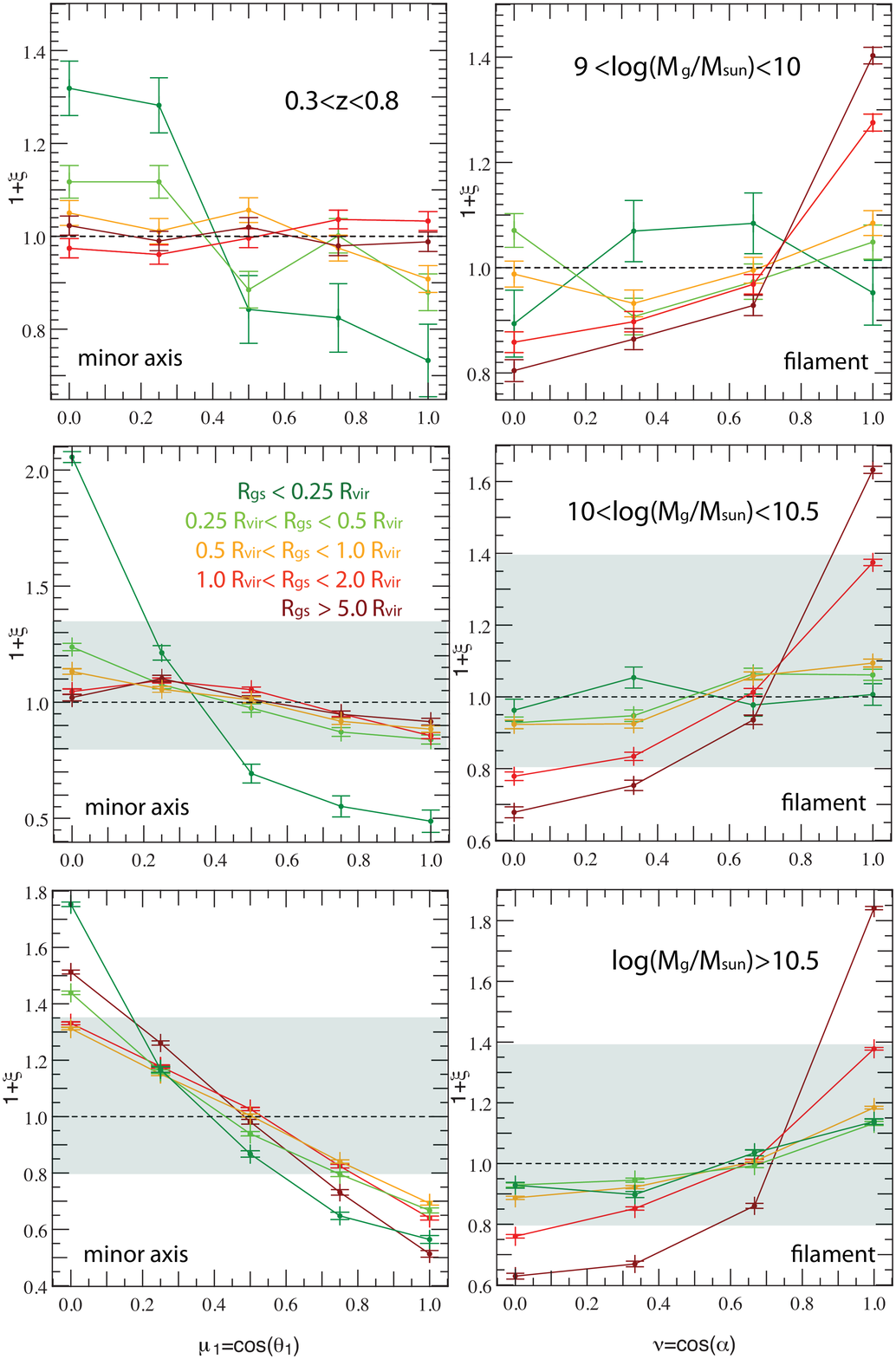}
  \caption{Same as Fig.~\ref{fig:mass0} where samples are binned in distance,  $R_{\rm gs}$, from satellite to central. This is plotted for three different central mass bins: $10^{9}< M_{\rm g} < 10^{10} \,\rm M_{\odot}$ (top panels), $10^{10} < M_{\rm g} < 10^{10.5} \,\rm M_{\odot}$ (middle panels) and $M_{\rm g} > 10^{10.5} \,\rm M_{\odot}$ (bottom panels). The amplitude of the signal for the lowest mass bin ($10^{9}< M_{\rm g} < 10^{10} \,\rm M_{\odot}$) is represented as a grey area in the plots obtained for higher mass bins.
Satellites close to the central galaxy tend to be distributed on the galactic plane with marginal alignment to the filament, while satellites in the outskirt of the host halo of the central are strongly aligned with the filament but the coplanarity with the central galaxy is weakened. Results are stacked for $0.3<z<0.8$.}
\label{fig:radius_mass}
\end{figure*}

	In order to further confirm this transition from filamentary to coplanar trend, Fig.~\ref{fig:radius_mass} reproduces the analysis of Fig.~\ref{fig:radius_tot}  for three different central mass bins: $10^{9} < M_{\rm g} < 10^{10} \, \rm M_{\odot}$, $10^{10} < M_{\rm g} < 10^{10.5} \, 
	\rm M_{\odot}$ and $M_{\rm g} > 10^{10.5} \, \rm M_{\odot}$. The evolution of both trends with respect to the mass of the central is consistent with the idea that more massive centrals should exert stronger torques and with their average mass-dependent orientation within the cosmic web.

	This confirms the general tendency already observed for all central galaxies with $M_{\rm g} >10^{10} \,\rm M_{\odot}$ in Fig.~\ref{fig:radius_tot}: satellites in the outskirts of the halo are aligned with the nearest filament. As they reach the inner parts of the halo they deviate from 
	the filament to align with the galactic plane of their central. Specific features in each mass bin tend to confirm this general scenario:
	\begin{itemize}
	\item For low mass centrals, the {\it coplanar trend} is significantly weaker than for centrals with $M_{\rm g} >10^{10} \,\rm M_{\odot}$ (for satellites within $0.25\ R_{\rm vir}$: $\xi=30\%$ at $\cos \theta_1=0$  instead of $\xi\simeq100\%$ for the two highest mass ranges). 
	This is consistent with a weaker torquing of lower mass centrals. However, the flip from the {\it filamentary trend} to the {\it coplanar trend} is more distinctive than for higher mass bins, which is directly related to the fact that those small mass centrals are under the swing 
	transition mass evaluated in \cite{duboisetal14}, and therefore are more likely to have a minor axis aligned with the filament, in which case both trends compete.
	\item The {\it filamentary trend} in the outskirts of the halo is mildly affected by the mass of the centrals. As expected, it undergoes a little increase and persists at shorter distances from the central for high mass centrals  for which {\it filamentary} and {\it coplanar} trends are more
	 likely to reinforce each other.
	\item For most massive central galaxies, the {\it coplanar trend} shows a general evolution very similar to that observed for lower masses but experiences a new increase, although somewhat limited, for satellites in the most outer parts of the halo ($R_{\rm gs}>2\ R_{\rm vir}$).
	 Satellites so distant can actually be satellites of a neighboring cluster. Thus, this trend is consistent with the Binggeli effect \citep{Binggeli82} that applies to massive clusters, which tend to align their rotation plane with that of their neighbours and is a hint of  
	the importance of the two-halo term discussed in further detail by \citet{Chisari15}.
	\end{itemize}

\section{Spin versus minor axis: effect on  satellite alignment}
\label{section:spin-minor}

\begin{figure*}
\center \includegraphics[width=1.5\columnwidth]{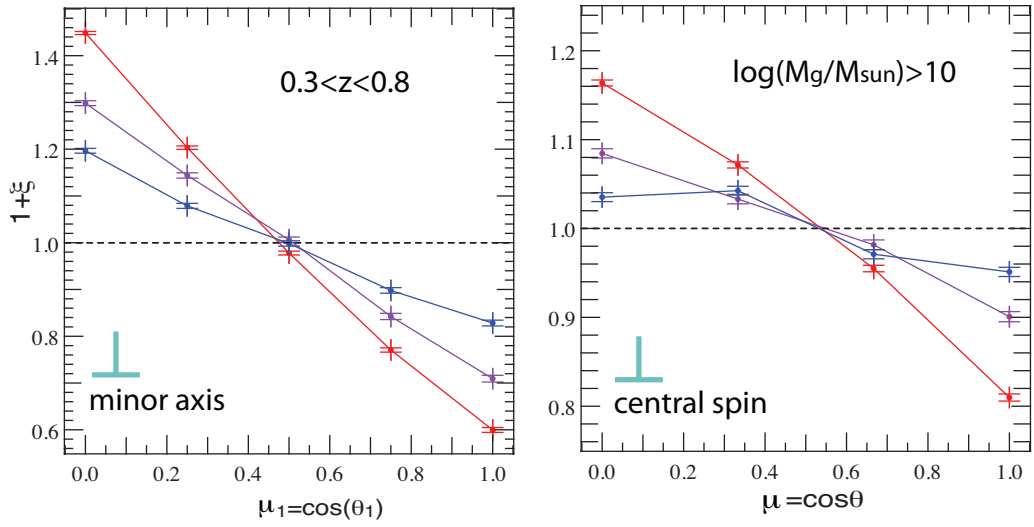}
 \caption{{\it Left panel}: PDF of $ \mu_{1}=\cos \theta_{1}$, the angle between the minor axis of the central galaxy and the vector separating it from its satellites, at $0.3<z<0.8$ and for different colour bins. {\it Right panel}: PDF of $ \mu=\cos \theta$, the angle between the spin of the central galaxy and the satellite separation vector. For massive red central galaxies, the satellites  tend to be distributed on the galactic plane. The amplitude of the signal is stronger when using the minor axis rather than the spin.}
\label{fig:colcut2}
\end{figure*}

Additionally, the right panel of Fig.~\ref{fig:colcut2} shows the PDF 
	of $ \mu=\cos \theta$, the angle between the spin of the central galaxy and the satellite separation vector. While replacing the alignment with the minor axis by the alignment with the galactic spin does not change our results qualitatively, one can see that the amplitude of 
	the spin signal is significantly lower than that of the axis signal, with the previously mentioned $54\%$ fraction of satellites within a solid angle sector around the midplane falling to less than $45\%$. 
	
	This is a general trend that we observed for several PDFs presented in this paper,
	 which suggests a significant impact of torquing from the central galaxy on the motion of satellites entering the halo. It also suggests that the spin can be significantly misaligned with both the minor axis of the central galaxy and the spin of the host halo. 
	The discrepancy between those two signals is highly dependent on the shape of the central galaxy, which can induce significant misalignments between the minor axis and the spin, especially for prolate structures. 

\end{document}